\documentclass[reprint, amsmath, amssymb,aps,prb,floatfix, eqsecnum, longbibliography,superscriptaddress]{revtex4-2}

\usepackage{amsmath,graphicx,natbib,bm,array,physics,caption,subcaption, hyperref}
\hypersetup{colorlinks = true,allcolors = blue }
\usepackage{bbm}
\usepackage{xcolor}
\usepackage[normalem]{ulem} 
\usepackage{mathtools}

\pdfoutput=1

\DeclareMathOperator{\sgn}{sgn}

\begin{document}

\newcommand{\pd}{\partial}
\newcommand{\beq}{\begin{equation}}
\newcommand{\eeq}{\end{equation}}
\newcommand{\bseq}{\begin{subequations}}
\newcommand{\eseq}{\end{subequations}}
\newcommand{\bpmat}{\begin{pmatrix}}
\newcommand{\epmat}{\end{pmatrix}}
\newcommand{\param}{\lambda}
\newcommand{\id}{\mathbb{I}}
\newcommand{\td}{T_{\text{drive}}}
\newcommand{\lloc}{L_{\text{loc}}}
\newcommand{\w}{\phi_{\text{max}}}
\newcommand{\dt}{\Delta t}
\newcommand{\z}{\mathbb{Z}}
\newcommand{\ident}{\mathbbm{1}}
\newcommand{\transl}{\hat{T}} 



\title{Universal localization-delocalization transition in chiral-symmetric Floquet drives}

\author{Adrian B. Culver}
\email[Present address: Center for Quantum Science and Engineering, Department of Electrical and Computer Engineering, University of California, Los Angeles, Los Angeles, California 90095, USA,\qquad]{adrianculver@g.ucla.edu}
\affiliation{Mani L. Bhaumik Institute for Theoretical Physics, Department of Physics and Astronomy, University of California Los Angeles, Los Angeles, California 90095, USA}

\author{Pratik Sathe}
\affiliation{Mani L. Bhaumik Institute for Theoretical Physics, Department of Physics and Astronomy, University of California Los Angeles, Los Angeles, California 90095, USA}
\affiliation{Theoretical Division (T-4), Los Alamos National Laboratory, Los Alamos, New Mexico 87545, USA}
\affiliation{Information Science \& Technology Institute, Los Alamos National Laboratory, Los Alamos, New Mexico 87545, USA}

\author{Albert Brown}
\affiliation{Mani L. Bhaumik Institute for Theoretical Physics, Department of Physics and Astronomy, University of California Los Angeles, Los Angeles, California 90095, USA}

\author{Fenner Harper}
\affiliation{Mani L. Bhaumik Institute for Theoretical Physics, Department of Physics and Astronomy, University of California Los Angeles, Los Angeles, California 90095, USA}

\author{Rahul Roy}
\email{rroy@physics.ucla.edu}
\affiliation{Mani L. Bhaumik Institute for Theoretical Physics, Department of Physics and Astronomy, University of California Los Angeles, Los Angeles, California 90095, USA}

\date{\today}

\begin{abstract}
    Periodically driven systems often exhibit behavior distinct from static systems.  In single-particle, static systems, any amount of disorder generically localizes all eigenstates in one dimension.  In contrast, we show that in topologically nontrivial, single-particle Floquet loop drives with chiral symmetry in one dimension, a localization-delocalization transition occurs as the time $t$ is varied within the driving period ($0 \le t \le \td$).
    We find that the time-dependent localization length $\lloc(t)$ diverges with a universal exponent as $t$ approaches the midpoint of the drive: $\lloc(t) \sim (t - \td/2)^{-\nu}$ with $\nu=2$.  We provide analytical and numerical evidence for the universality of this exponent within the AIII symmetry class.
\end{abstract}

\maketitle

\setcounter{footnote}{0} 
\footnotetext[0]{Here and throughout, our Hamiltonians and unitary time evolutions refer to single-particle matrices, not to the corresponding multiparticle operators that act on Fock space.  Thus, $\mathcal{C}$ is unitary rather than antiunitary \cite{ChiuClassification2016}.}

\section{Introduction}\label{sec:Introduction}

The interplay of disorder and topology can produce localization-delocalization transitions in both undriven (static) and driven (Floquet) systems.  This phenomenon is particularly prominent in low-dimensional systems, in which the tendency of disorder to lead to Anderson localization is strongest.

Let us recall an illustrative example from static systems.  In the integer quantum Hall effect plateau transition \cite{HuckesteinScaling1995}, there is a conflict between disorder, which favors localization, and a nontrivial value of a topological index, which forbids complete localization.  Indeed, the nonzero Chern number implies that each Landau band must include a delocalized state (at some energy $E_c$), while the remaining states are generically localized by disorder \cite{HalperinQuantized1982}.  As the energy $E$ crosses the transition point $E_c$, the localization length $\lloc$ diverges on both sides as $\lloc \sim 1/|E-E_c|^\nu$ with a universal exponent $\nu$ whose precise value remains controversial \cite{IppolitiDimensional2020}.

In the periodic table of static topological insulators and superconductors \cite{KitaevPeriodic2009}, localization-delocalization transitions have been studied in various Altland-Zirnbauer symmetry classes \cite{KagalovskyUniversal2008, EversAnderson2008, SongAIII2014, SongEffect2014}.  In the case of the chiral-symmetric class (AIII) in one spatial dimension, a localization-delocalization transition at zero energy has been shown to accompany a topological phase transition \cite{Mondragon-ShemTopological2014} (see also Refs. \cite{SongEffect2014, XiaoAnisotropic2023} for the 3D case).  Furthermore, broader theoretical approaches to localization-delocalization transitions have been proposed, including a nonperturbative transfer-matrix method for quasi-one-dimensional systems in several symmetry classes  \cite{AltlandTopology2015} and a random Dirac Hamiltonian method for the full periodic table \cite{MorimotoAnderson2015}.

Localization-delocalization transitions have also been studied in Floquet systems.  Examples include the Floquet topological Anderson insulator \cite{TitumDisorderInduced2015, RoyDisordered2016, TitumDisorderinduced2017a}, discrete-time quantum walks \cite{ObuseTopological2011, KitagawaTopological2012, RakovszkyLocalization2015, VakulchykAnderson2017, DerevyankoAnderson2018} (which are closely related to Floquet systems and which have also been proposed as a means to realize static topological phases \cite{KitagawaExploring2010}), and an effective Hamiltonian theory for disorder-induced transitions from topologically nontrivial to trivial phases \cite{ShtankoStability2018}.  Let us also mention Ref. \cite{WautersLocalization2019}, which finds, in the one-dimensional driven Rice-Mele model, a transition similar to the integer quantum Hall plateau transition.  Indeed, Ref. \cite{WautersLocalization2019} finds a localization-delocalization transition from a topologically nontrivial phase to a trivial phase at a critical value of disorder strength $W=W_c$; furthermore, the localization length of the Floquet states diverges as $\lloc \sim (W-W_c)^{-\beta}$ with $\beta \simeq 2$.

When prior work on localization-delocalization phenomena in the Floquet (or quantum walk) case has considered symmetries \cite{ObuseTopological2011, KitagawaTopological2012, RakovszkyLocalization2015, GannotEffects2015, VakulchykAnderson2017}, the focus has generally been on the symmetries of the Floquet Hamiltonian $H_F$, which determines the time evolution operator for a single driving period.  For instance, chiral symmetry $\mathcal{C}$ imposes the requirement $\mathcal{C} H_F \mathcal{C}^{-1}= - H_F$ \cite{Note0}.  However, a full account of topological possibilities in Floquet systems requires consideration of further details of the drive beyond $H_F$; indeed, a topologically nontrivial drive can occur even if $H_F$ acts as the identity operator in the bulk (see Ref. \cite{HarperTopology2020} and references therein).  In the periodic table of Floquet topological insulators \cite{RoyPeriodic2017}, the symmetries used in the classification relate the time-dependent Hamiltonian $H(t)$ to itself at another time, e.g., $\mathcal{C} H(t) \mathcal{C}^{-1} = -H(\td-t)$ for chiral symmetry (where $\td$ is the driving period).  We therefore propose to study, in one particular entry in the periodic table, the interplay of topology and of disorder that respects the appropriate time-dependent symmetry constraint.

The main result of this paper is a universal localization-delocalization transition for disordered, topologically nontrivial, single-particle Floquet drives in the chiral-symmetric class in one spatial dimension.  We consider the time-dependent localization length $\lloc(t)$ of a time evolution operator that represents the intrinsically dynamical part of the time evolution operator for the drive (see next paragraph), and we find that $\lloc(t)$ diverges with a universal exponent at a particular $t$ within the driving period.  As in the integer quantum Hall plateau transition, we have a topological obstruction to localization at a particular value of some parameter, which then leads to a localization-delocalization transition as that parameter is varied.  Unlike the quantum Hall case or the one-dimensional class AIII static case (Ref. \cite{Mondragon-ShemTopological2014} mentioned above), here the parameter is time and the transition occurs throughout the (quasienergy) spectrum.  Also, the transition we find does not seem to be a topological \emph{phase} transition; see Sec. \ref{sec:Universal localization-delocalization transition} below for more discussion of this point.

To give a more detailed statement of our result, we recall first that any drive can be decomposed (uniquely up to homotopy) into two component drives: a static drive and an intrinsically dynamical ``loop'' drive $U_\text{loop}(t)$ ($0\le t \le \td$), where the latter by definition is equal to the identity at both $t=0$ and $\td$ \cite{RoyPeriodic2017}.  The decomposition can be done such that the loop component $U_\text{loop}$ inherits the symmetries of the originally given drive, in this case chiral symmetry.  Our focus is entirely on the loop component $U_\text{loop}$, which we emphasize contains the intrinsically dynamical possibilities for topology.  Assuming $U_\text{loop}$ to realize the simplest possibility for topological nontriviality---a value of $1$ for the appropriate topological index \cite{LiuChiral2018}---we study the time-dependent localization length $\lloc(t)$ of $U_\text{loop}(t)$   \footnote{In particular, $\lloc(t)$ is the localization length for the time-periodic Floquet states of the operator $U_\text{loop}(t)$ considered as if $t$ were the full period.}.  We argue that the chiral symmetry and topological nontriviality require delocalization at the midpoint of the drive ($t=\td/2$).  Since we are considering one spatial dimension, we can expect complete localization at all other times $t$, since there is no topological protection there.  Our main claim is that there is a universal exponent for the transition that occurs as $t$ approaches $\td/2$ from either side: $\lloc(t) \sim |t-\td/2|^{-\nu}$ with $\nu=2$.  The localization length generally depends on the quasienergy and on the disorder, but we claim that generically these only appear in the prefactor and not the exponent.

Our work is not directly motivated by experiment, and indeed it may be expected that both the loop decomposition and the requirement for the disorder to obey a time-dependent symmetry would present challenges to detecting the diverging localization length in a real system.  One possibility for experimental realization is the close connection between Floquet drives and discrete-time quantum walks.  Also, the transition we find here could serve as a theoretical toy model for other transitions.

The particular disordered Floquet drive that we use as an example for our calculations has been studied in an equivalent form in the context of discrete-time quantum walks \cite{VakulchykAnderson2017, DerevyankoAnderson2018}, as we discuss in more detail in the main text.  Our claim of the universal transition for a class of drives goes beyond prior work to our knowledge.  Preliminary work toward the results of this paper was reported by some of the present authors in Refs. \cite{BrownGeometric2019,SatheAspects2023}.

The paper is organized as follows.  In Sec. \ref{sec:Disordered, chiral-symmetric drives in one dimension}, we review the topological classification of noninteracting Floquet drives and the symmetry class AIII that is our main focus.  We then state our main result, clarify the role of chiral-symmetric disorder, and discuss a possible wider scope for the universal exponent (as well as some ``fine-tuned'' exceptions). In Sec. \ref{sec:Example drives with the universal exponent}, we present analytical and numerical calculations illustrating the universal exponent in several examples of class AIII drives.  In Sec. \ref{sec:Scattering argument for the universal exponent}, we present a more general argument for the universal exponent.  We conclude in Sec. \ref{sec:Conclusion} with some ideas for investigating many-body localization in class AIII drives that include interactions.

\section{Disordered, chiral-symmetric drives in one dimension} \label{sec:Disordered, chiral-symmetric drives in one dimension}

\subsection{Preliminary discussion}\label{sec:Preliminary discussion}

Periodically driven (Floquet) systems are described by a time-periodic Hamiltonian $\mathcal{H}(t)$ with a period $\td$, so that $\mathcal{H}(t+\td) = \mathcal{H}(t)$.  Throughout, we consider real-space local Hamiltonians, i.e., the matrix elements in position space must be upper bounded by some decaying exponential.  The unitary time evolution operator $\mathcal{U}(t)$ is given by the usual time-ordered expression $\mathcal{U}(t) = T \exp\left[ - i \int_0^t dt'\ \mathcal{H}(t') \right]$.  Note that $\mathcal{U}(t)$ need not be periodic, even though $\mathcal{H}(t)$ is.

Noninteracting Floquet topological insulators have been classified into a periodic table \cite{RoyPeriodic2017} (see Ref. \cite{HarperTopology2020} for a review) using the Altland–Zirnbauer symmetry classification.  In particular, Floquet drives are separated into ten symmetry classes based on the presence or absence of time-reversal, particle-hole, and chiral symmetry, and the full topological classification depends on the symmetry class and the number of spatial dimensions.  We focus on 1D class AIII systems, i.e., those with chiral symmetry only. This is the simplest case in which a topologically nontrivial Floquet phase can occur.

By definition, class AIII drives satisfy
\beq
    \mathcal{C} \mathcal{H}(t) \mathcal{C}^{-1} = - \mathcal{H}(\td-t),\label{eq:chiral symmetry mathcalH(t)}
\eeq
where $\mathcal{C}$ denotes the chiral symmetry operator, which satisfies $\mathcal{C}^2 = \mathcal{C} \mathcal{C}^\dagger = \mathcal{C}^\dagger \mathcal{C} = \mathbbm{1}$ and $\text{Tr}\  \mathcal{C}=0$.  The corresponding unitary evolution $\mathcal{U}(t)$ inherits a symmetry property (see Appendix A of Ref. \cite{RoyPeriodic2017}):
\beq
    \mathcal{C} \mathcal{U}(t)\mathcal{C}^{-1} = \mathcal{U}(\td - t) \mathcal{U}^\dagger (\td).\label{eq:chiral symmetry mathcalU(t)}
\eeq
We suppose that our Floquet drive acts on a one-dimensional lattice with basis states $\ket{n,\alpha,s}$, where $n$ is a site index taking on integer values, $\alpha=A$ or $B$ is a sublattice index on which the chiral symmetry acts, and $s$ stands for any other on-site degrees of freedom.  The chiral symmetry operator can be written as
\beq
    \mathcal{C} = \tau^z,\label{eq:chiral symmetry as tauz}
\eeq
where $\tau^z$ is the Pauli matrix acting on the sublattice index (and acting as the identity on the other indices).

Our focus in this work is on topological properties that are uniquely dynamical, i.e., not simply inherited from a static Hamiltonian.  For this reason, we find it convenient to use the technique introduced in Ref. \cite{RoyPeriodic2017} (see also Ref. \cite{LiuChiral2018}) of decomposing a generic drive into a particular combination of time evolution by a static Hamiltonian and time evolution by a loop drive.

To present the loop decomposition, we first recall the definition of composition of two drives.  Consider two chiral-symmetric drives $\mathcal{U}_1(t)$ and $\mathcal{U}_2(t)$ ($0\le t \le \td$) which are generated by time-dependent Hamiltonians $\mathcal{H}_1(t)$ and $\mathcal{H}_2(t)$.  The composition operation \cite{RoyPeriodic2017}, denoted $*$, produces a new chiral-symmetric drive $\mathcal{U}(t)$ with the same period; suppressing $t$, we write $\mathcal{U}\equiv \mathcal{U}_1 * \mathcal{U}_2$.  By definition, the drive $\mathcal{U}$ is generated by a time-dependent Hamiltonian $\mathcal{H}\equiv \mathcal{H}_1 * \mathcal{H}_2$ given by
\beq
    \mathcal{H}(t) = 
    \begin{cases}
    2\mathcal{H}_2(2t) & 0 \le t < T_\text{drive}/4 \\
    2\mathcal{H}_1(2t - T_\text{drive}/2) & T_\text{drive}/4  \le t \le 3T_\text{drive}/4 \\
    2\mathcal{H}_2(2t-\td) & 3T_\text{drive}/4  < t \le T_\text{drive}.
    \end{cases}
\eeq
It is readily verified that $\mathcal{H}(t)$ satisfies the chiral symmetry property \eqref{eq:chiral symmetry mathcalH(t)}.

A loop drive is one that begins and ends at the identity when the system has periodic boundary conditions (or is infinite): $U_\text{loop}(t=0) = U_\text{loop}(t=\td)= 1$.  A static drive is one generated by some time-independent Hamiltonian $H_\text{static}$:
\beq
    U_\text{static}(t) = e^{-i H_\text{static} t}.\label{eq:Ustatic}
\eeq
As shown in Ref. \cite{RoyPeriodic2017}, an arbitrary chiral-symmetric drive $\mathcal{U}(t)$ is homotopic to the composition of some chiral-symmetric loop drive $U_\text{loop}(t)$ and some static drive $U_\text{static}(t)$:
\beq
    \mathcal{U} \approx U_\text{loop} * U_\text{static},\label{eq:loop decomposition}
\eeq
where $U_\text{loop}$ and $U_\text{static}$ are unique up to homotopic equivalence.  We will focus on a canonical choice defined by taking $U_\text{loop}(t)$ to be generated by $H_\text{loop}(t)$ and $U_\text{static}(t)$ to be as in Eq. \eqref{eq:Ustatic}, with
\bseq
\begin{align}
    H_\text{loop} &= \mathcal{H}*(-H_F),\\
    H_\text{static} &= H_F,
\end{align}
\eseq
where $H_F$ is the Floquet Hamiltonian for the drive [$\mathcal{U}(\td) = e^{-i H_F \td}$].  Since our interest is in $t$ near the midpoint, let us note here that the above equations readily yield (for $\td/4 \le t \le 3\td/4$)
\beq
    U_\text{loop}(t) = \mathcal{U}(2t - \td /2 ) e^{-iH_F \td/2}.\label{eq:Uloop(t) near midpoint}
\eeq

We recall here that the loop decomposition  requires the endpoint unitary $\mathcal{U}(\td)$ to have, in the translation-invariant case, a spectral gap at quasienergy $\epsilon=\pi$.  We are interested in the disordered case, so we can expect that $\mathcal{U}(\td)$ has only localized states and no spectral gap.  However, the entire spectrum is then a mobility gap, and a mobility gap in particular at $\epsilon=\pi$ should suffice for doing the loop decomposition (see Refs. \cite{RoyPeriodic2017,GrafBulk2018}). 

The decomposition \eqref{eq:loop decomposition} is essential to this work, as our focus throughout is on the localization properties of the loop component of the given drive $\mathcal{U}(t)$.  From now on, we simplify the notation by writing the loop component as $U_\text{loop}(t) \equiv U(t)$.

We are interested in drives for which the loop component is topologically nontrivial.  To state this condition precisely, we now review the chiral flow, the topological index introduced in Ref. \cite{LiuChiral2018} for the loop component of class AIII drives in one dimension.  We first recall the flow index (the topological index for unitary operators without symmetry in one dimension \cite{KitaevAnyons2006}), which is used in the definition of the chiral flow.

Given a unitary matrix $U$ acting on an infinite one-dimensional lattice, with matrix elements $U_{na,n'a'}$ (where $n$ is the lattice site index and $a$ labels any on-site degrees of freedom), the flow index of $U$ is defined as
\beq
    \nu(U) = \sum_{\substack{n\geq n_0 \\ n'<n_0} }\sum_{a,a'}\left( |U_{na,n'a'}|^2-|U_{n'a', na}|^2 \right),
\eeq
where the site $n_0$ is arbitrary [and $\nu(U)$ is in fact independent of $n_0$].  The flow index is quantized to integer values and may be understood as the net current passing through any site $n_0$.  For example, the operator that translates by a certain integer has a flow index equal to that integer (e.g., the mapping $\ket{n}\to\ket{n+1}$ has a flow index of $1$).

To apply the flow index to a class AIII drive, we note that the loop component of the drive inherits the same chiral symmetry property \eqref{eq:chiral symmetry mathcalU(t)}, which simplifies due to the loop property to
\beq
    \mathcal{C} U(t)\mathcal{C}^{-1} = U(\td - t).\label{eq:chiral symmetry U(t)}
\eeq
Due to this symmetry, the midpoint of the loop part of the drive is a special point; in particular, we have the following useful relation \cite{LiuChiral2018}:
\beq
    \mathcal{C} U(\td/2) \mathcal{C}^{-1} = U(\td/2).\label{eq:chiral symmetry midpoint}
\eeq
Recalling the form \eqref{eq:chiral symmetry as tauz} of the chiral symmetry operator, we see that Eq. \eqref{eq:chiral symmetry midpoint} constrains the time evolution at the midpoint to the block diagonal form
\beq
    U(\td/2) = 
    \bpmat
        U_{AA} & 0 \\
        0 & U_{BB}
    \epmat.\label{eq:Umidpoint as UAA+UBB}
\eeq
We may then consider the flow index of the two components $U_{AA}$ and $U_{BB}$.  Since the time evolution operator at the midpoint is generated (by assumption) by a local Hamiltonian, it follows that $\nu[U(\td/2)]=0$ \cite{GrossIndex2012,RanardConverse2022}.  From the additivity of the flow index \cite{GrossIndex2012}, we then see that the flow indices of the two components sum to zero: $\nu(U_{AA})+\nu(U_{BB})=0$.  Either of the two may be taken as the definition of the chiral flow of the loop drive $U(t)$.  We follow the convention that the chiral flow is the flow index of the $A$ component \cite{LiuChiral2018}:
\beq
    \nu_{\text{chiral}}[\{U(t)\}] = \nu[U_{AA}].\label{eq:chiral flow def}
\eeq

We are interested in topologically nontrivial drives, so we assume that the chiral flow is nonzero.  More specifically, our numerics and analytical arguments focus on the case of $\nu_{\text{chiral}}[\{U(t)\}]=1$.  Let us note here that a drive with unit chiral flow can be generated by a local Hamiltonian, as shown by the explicit model introduced in Ref. \cite{LiuChiral2018} (and reviewed below in Sec. \ref{sec:Example drives with the universal exponent}); in contrast, no local Hamiltonian can generate a unitary with nonzero flow index in one dimension \cite{GrossIndex2012,RanardConverse2022}.

\subsection{Universal localization-delocalization transition}\label{sec:Universal localization-delocalization transition}
It is well-known that in one dimension, the eigenstates of a disordered, static Hamiltonian are generically localized.  We can expect the same to be true for eigenstates of $U(t)$ for a generic value of the time $t$.  However, we argue that the midpoint of the drive $(t=\td/2)$ is topologically protected from localization by our assumption that the chiral flow is non-vanishing.  (We do not yet make the more specific assumption that the chiral flow is equal to one.)

In particular, we present two independent arguments for the following claim: at almost any quasienergy $\epsilon$ in the spectrum of $U(\td/2)$, there must exist a delocalized eigenstate.  We have checked this claim, sometimes analytically and sometimes numerically, in a variety of cases (see below).  A related analysis will be presented elsewhere \cite{SatheTopologically2025}.

Due to the diagonal structure of Eq. \eqref{eq:Umidpoint as UAA+UBB}, our claim reduces to the statement that a local unitary operator $U'$ (which represents either $U_{AA}$ or $U_{BB}$) with non-zero flow index must have delocalized states throughout its quasienergy spectrum.  To sketch our first argument, we start by recalling that all $U'$ with the same nonvanishing flow index are homotopic to a translation operator (which translates by a number of sites equal to the value of the flow index) \cite{GrossIndex2012}.  A translation operator indeed has delocalized eigenstates throughout its spectrum, and furthermore these states are chiral, so they are perturbatively robust, similar to the edge states in a quantum Hall system.

For our second argument, we appeal to the general expectation that mobility edges are not found in one dimension.  Here we ignore the known exceptions of correlated disorder and quasiperiodic systems \cite{LiuAnomalous2022} because our focus is on the generic case.  Thus, it should suffice to show that $U'$ must have a delocalized state at \emph{some} quasienergy.  To show this, we write $U'$ in its eigenbasis:
\beq
    U' = \sum_\epsilon e^{-i \epsilon \td/2} \ket{\Psi_\epsilon}\bra{\Psi_\epsilon}.\label{eq:U eigenmode expansion}
\eeq
If every eigenvector $\ket{\Psi_\epsilon}$ is localized, then we can take the logarithm of Eq. \eqref{eq:U eigenmode expansion} to obtain a \emph{local} Hamiltonian $H_F$ that generates $U'$ (via $U'=e^{-i H_F \td/2}$).  But a unitary operator that is generated by a local Hamiltonian must have zero flow index \cite{GrossIndex2012,RanardConverse2022}, in contradiction to our starting assumption.  Thus, we conclude that some eigenvector $\ket{\Psi_\epsilon}$ must be delocalized (hence one expects all eigenvectors to be delocalized).

To set up the statement of our main claim, let us write the eigenstate equation for $U(t)$ explicitly:
\beq
    U(t) \ket{\Psi_\epsilon (t)} = e^{- i\epsilon t} \ket{\Psi_\epsilon(t)},\label{eq:eigenstate eq}
\eeq
where $\epsilon$ is the quasienergy and where the $t$ dependence in $\ket{\Psi_\epsilon(t)}$ is a label associating this state to $U(t)$ (\emph{not} an indication of time evolution by the time-dependent Schrodinger equation).  Since we are considering an infinitely large, disordered system, the quasienergy spectrum of $U(t)$ is gapless for all $t$.

Let us emphasize that we are focusing on the time evolution operator $U(t)$ with arbitrary $t$, not necessarily the full period ($t=\td$). One perspective on this is to consider a discrete-time process in which $U(t)$ is the operator for implementing a single time step; in other words, any given $t$ is considered to be the full period.  The properties of the eigenstates of $U(t)$ would have implications for this discrete-time process.  It is in the same spirit that we use the term ``quasienergy'' in reference to the spectrum of $U(t)$.

We write the localization length of $\ket{\Psi_\epsilon(t)}$ as $\lloc(t)$, suppressing the dependence on $\epsilon$.  According to our arguments above, $\lloc(t)$ is infinite for $t=\td/2$ and finite otherwise.  For a drive with $\nu_\text{chiral}[\{U(t)\}]=1$, we claim that the localization length diverges with a universal exponent as $t$ approaches the midpoint:
\beq
    \lloc(t) \sim \frac{1}{\left( t- \td/2\right)^\nu} \qquad (\nu=2).\label{eq:main claim}
\eeq
Equation \eqref{eq:main claim} is the main claim of this paper.  The constant of proportionality is non-universal and generally depends on the quasienergy; however, the exponent $\nu=2$ holds throughout the quasienergy spectrum.  We emphasize that the localization length we are considering corresponds to the instantaneous eigenstates of the loop part $U(t)\equiv U_\text{loop}(t)$ of the original drive.

We argue that the exponent of $\nu=2$ is generic for loop drives in class AIII with $\nu_\text{chiral}[\{U(t)\}]=1$, albeit with certain fine-tuned exceptions that we discuss below in Sec. \ref{sec:Exceptions to the universal exponent, and further generality}.

Before presenting more detailed arguments, let us mention that there is a simple way to obtain the exponent $\nu=2$, granting the assumption (which our calculations below support) that the Lyapunov exponent $1/\lloc(t)$ is an analytic function at $t=\td/2$.  Due to the topological argument we discussed above, the expansion of $1/\lloc(t)$ about the point $t=\td/2$ can have no zeroth order term; then there can be no linear term because the Lyapunov exponent cannot be negative.  Thus, this simple argument yields the series starting generically at quadratic order, which is precisely \eqref{eq:main claim}.  Our more detailed argument in Sec. \ref{sec:Scattering argument for the universal exponent} can be roughly summarized as showing that, under some simplifying assumptions, the Lyapunov exponent is given at the leading order by a second-order polynomial in variables that start at linear order in $t-\td/2$. 

Let us also comment here on the similarities and differences between our work and the integer quantum Hall effect plateau transition.  In both cases, we have a varying parameter (time $t$ or energy $E$) and a topological index (the chiral flow or the Chern number) that is associated with a range of values of the parameter (the driving period or the bandwidth).  The eigenstates are localized at all values of the parameter except for some critical value ($\td/2$ or $E_c$), where delocalization must occur due to the index being nonzero.  A localization-delocalization transition therefore occurs as the parameter crosses the critical value.  One crucial difference is that the quantum Hall case has a topological phase transition, while our case, as far as we know, does not.  The quantum Hall case is a static problem, so a topological phase can be defined by the index of the ground state, and thus a topological phase transition occurs as the Fermi energy crosses $E_c$ (since the extended state carries the Chern number).  In a Floquet problem, though, it is not clear what could play the role of the ground state or Fermi energy.  Another difference is that there is a noninteger delocalization exponent in the quantum Hall case (in particular, the Lyapunov exponent is nonanalytic at $E=E_c$); in our case, we have a scattering argument that yields analyticity at $t=\td/2$ and thus an integer exponent. 

\subsection{Chiral-symmetric disorder in loop drives}\label{sec:Chiral-symmetric disorder in loop drives}
In this section, we clarify the role of disorder in our setup.  Our main claim \eqref{eq:main claim} is meant to apply to any \emph{particular} drive, i.e., the thermodynamic limit of a single disorder realization (as long as the realization is suitably generic).  For calculations, we find it convenient to use the idea of self-averaging from the theory of disordered systems, which refers to the fact that for appropriately chosen physical quantities, the relative fluctuations over an ensemble of disorder realizations vanish in the thermodynamic limit.  For these quantities, the value in any particular system can also be obtained by averaging over an ensemble of disorder realizations.  Note that the localization length that we have discussed above is really the thermodynamic limit of a size-dependent localization length and furthermore, this quantity may be expected to be self-averaging.

Our basic assumption is that the given drive $\mathcal{U}(t)$ is in some sense a disordering of an underlying translationally invariant (``clean'') drive $\mathcal{U}_\text{clean}(t)$.  According to self-averaging, we could determine properties of a particular, generic $\mathcal{U}(t)$ by averaging over an ensemble of disorder realizations (each chiral-symmetric) of $\mathcal{U}_\text{clean}(t)$; however, this is inconvenient for calculations because we would then need to do the decomposition \eqref{eq:loop decomposition} on each member of the ensemble.

We instead proceed as follows: we do the decomposition \eqref{eq:loop decomposition} on
$\mathcal{U}(t)$ and on $\mathcal{U}_\text{clean}(t)$ to yield loop drives $U(t)$ and $U_\text{clean}(t)$, respectively.  We then regard $U(t)$, which is a particular drive, as a member of an ensemble of disorderings of $U_\text{clean}(t)$.  Each member of the ensemble is a chiral-symmetric loop drive.  By considering a range of possible disorder distributions in such an ensemble, we expect that properties of a generic starting drive $\mathcal{U}(t)$ are being probed.

We now present, for use in our subsequent calculations in specific example models, one particular way of adding disorder to a given translationally invariant, chiral-symmetric loop drive $U_\text{clean}(t)$.  We provide this construction because it is not immediately obvious how to include disorder that respects the required properties, even if one is given the explicit Hamiltonian $H_\text{clean}(t)$ that generates $U_\text{clean}(t)$; disordering the parameters of $H_\text{clean}(t)$ generally results in a drive that fails to satisfy the chiral symmetry requirement, the loop condition, or both.  After presenting this particular construction for including disorder, we discuss some general properties of a disordered drive that hold regardless of the precise way the disorder is included.

Our construction takes as given, in addition to the chiral-symmetric Hamiltonian $H_\text{clean}(t)$ with period $\td$ that generates $U_\text{clean}(t)$, a static disordered Hamiltonian $H_d$ that commutes with the chiral symmetry operator:
\beq
    \mathcal{C}^{-1} H_d \mathcal{C} = H_d.\label{eq:Hd commutes with C}
\eeq
Note that this condition is equivalent to $H_d$ being block diagonal in the A-B basis.  We further assume that $H_d\to 0$ in the limit of no disorder.

We choose an arbitrary constant time $T_0>0$ and define a new drive $H(t)$ with a period of $\td + T_0$:
\begin{widetext}
\beq
    H (t) =
    \begin{cases}
        H_d & 0 < t < T_0 / 2\\
        H_\text{clean}(t - T_0/2) & T_0 / 2 \le  t \le \td+ T_0 / 2\\
        - H_d & T_0 / 2 + \td < t \le \td + T_0, 
    \end{cases}\label{eq:disordering_a_clean_drive}
\eeq
with $H (t + \td + T_0) = H(t)$.  The time evolution operator from $t=0$ to any $t \in [0, \td + T_0]$, expressed in terms of $U_\text{clean}(t)$ and $H_d$, is
\bseq
\begin{align}
    U (t) &\equiv T \exp \left( -i \int_{0}^t dt'\ H(t') \right) \\
    &= 
    \begin{cases}
        e^{-i H_d t} & 0 < t < T_0/2\\
        U_\text{clean}(t - T_0/2) e^{-i H_d T_0/2} & T_0/2 \le t \le T_0/2 + \td\\
        e^{i H_d (t-\td - T_0)}  & T_0/2 + \td < t \le \td + T_0,
    \end{cases}\label{eq:disordered unitary with T0}
\end{align}
\eseq
\end{widetext}
where we have used the loop condition $U_\text{clean}(\td) = 1$.  It is readily verified that $U(t)$ is a chiral-symmetric loop drive; in particular, $H(t)$ satisfies \eqref{eq:chiral symmetry mathcalH(t)} (with the period increased from $\td$ to $\td + T_0$) and $U(\td + T_0) = 1$.

For convenience, we now take a limit so the disordered drive has the same period as the clean drive.  To do this, we send $T_0\to 0$ with $H_d$ scaling with $T_0$ in such a way that $H_d T_0$ remains fixed.  We thus obtain a disordered unitary operator
\beq
    U_d =\lim_{\substack{T_0\to 0 \\ H_d T_0\text{ fixed}}} e^{-i H_d T_0/2},\label{eq:Ud general}
\eeq
and we may write the disordered drive with a period of $\td$:
\beq
    U(t) = U_\text{clean}(t) U_d \qquad (0\le t < \td).\label{eq:disordered unitary one-sided}
\eeq
Equation \eqref{eq:disordered unitary one-sided} is to be understood as shorthand for the limit of Eq. \eqref{eq:disordered unitary with T0}.  In particular, it is understood that if $t$ is set exactly to $\td$, then an additional factor of $U_d^\dagger$ appears multiplying $U_\text{clean}(\td)=1 $ on the left, which restores the loop condition [$U(\td)=1$].  Then the chiral symmetry relation \eqref{eq:chiral symmetry mathcalH(t)} holds due to $\mathcal{C}U_d \mathcal{C}^{-1}=U_d$, which follows from the assumption that $H_d$ commutes with $\mathcal{C}$. 

We next present an alternate approach in which the disordered drive is not explicitly constructed, but instead expanded about the midpoint of the drive (which is the regime of interest for the universal exponent).  This approach clarifies the assumption we make that the drive is generic; indeed, we show that if a certain fine-tuned condition is satisfied, then the localization length may diverge with an exponent greater than $2$.

A general drive may be expanded about the midpoint ($t=\td/2$) in powers of $\dt$:
\begin{multline}
    U(\td /2 + \dt) = \left[ 1 - i H(\td/2)\dt +O((\dt)^2) \right]\\
    \times U(\td /2).\label{eq:generic expansion of U(t) near midpoint}
\end{multline}
If $H(t)$ is piecewise smooth with a discontinuity at $\td/2$, then $H(\td/2)$ here is replaced by its right (left) limit as $t\to \td/2$ when $\dt$ is positive (negative).

The expansion \eqref{eq:generic expansion of U(t) near midpoint} provides an alternate way to probe the ensemble of allowed disorderings of a given translation-invariant loop drive $U_\text{clean}(t)$.  The disordered drive $U(t)$ must at least respect the chiral symmetry and the loop property, in addition to reducing to $U_\text{clean}(t)$ in the absence of disorder.  It may be difficult to find explicit examples [aside from the construction \eqref{eq:disordered unitary one-sided} that we have already provided] of $U(t)$ that satisfy these properties; however, we know that \emph{any} valid $U(t)$ has some $U(\td/2)$ and $H(\td/2)$ [with $U(\td/2)$ satisfying the symmetry relation \eqref{eq:chiral symmetry midpoint}], which reduce to the corresponding terms for $U_\text{clean}(t)$ in the absence of disorder.  We expect, then, that for $t$ near the midpoint, we can probe the ensemble of allowed disorderings of the drive $U_\text{clean}(t)$ by specifying various disordered operators that act on the same Hilbert space as $U_\text{clean}(t)$ and then assuming these operators to be $U(\td/2)$ and $H(\td/2)$ of some $U(t)$.  We can then use the expansion \eqref{eq:generic expansion of U(t) near midpoint} without knowing the full form of $U(t)$.

We do not pursue the problem of completing the drive, i.e., constructing a chiral-symmetric loop drive $U(t)$ for all $t$ that recovers the given $U(\td/2)$ and $H(\td/2)$.  However, we expect that some chiral-symmetric loop drive completion can always be found.

\subsection{Exceptions to the universal exponent, and further generality}\label{sec:Exceptions to the universal exponent, and further generality}
Here we identify a necessary condition for the exponent to be $\nu=2$.  We argue that this condition should hold for a generic drive (with possible fine-tuned exceptions).  This discussion naturally leads us to a wider setting, not necessarily connected to chiral-symmetric drives, in which we claim that our main result \eqref{eq:main claim} holds.

Note first that we can decompose any operator $\mathcal{O}$ into parts that commute and anticommute with the chiral symmetry operator:
\beq
    \mathcal{O} = \mathcal{O}_- +\mathcal{O}_+,
\eeq
where $[\mathcal{O}_-,\mathcal{C} ] = \{ \mathcal{O}_+, \mathcal{C} \} = 0$.  Applying this to $\mathcal{O}=H(t)$, we see that the anticommuting part of the chiral symmetry equation \eqref{eq:chiral symmetry mathcalH(t)} for $t=\td/2 + \dt$ yields
\beq
    H_+(\td/2 + \dt) = H_+(\td/2 - \dt), 
\eeq
where $\dt$ is arbitrary.  In particular, $H_+(t)$ is continuous at $t=\td/2$, so we can consider $H_+(\td/2)$ without needing to distinguish the left and right limits.

We will assume, as part of the definition of a drive being generic, that the anticommuting part of the Hamiltonian at the midpoint is nonvanishing:
\beq
    H_+(\td/2)\ne 0.\label{eq:Hplus non-zero}
\eeq
Note that there is no symmetry that requires this term to vanish.  If instead $H_+(\td/2)=0$, then the $\dt$ correction term in the expansion \eqref{eq:generic expansion of U(t) near midpoint} commutes with $\mathcal{C}$; then, $U(\td/2 + \dt)$ remains block-diagonal in the $A$-$B$ basis up to corrections of order $(\dt)^2$. The block-diagonal structure preserves the topological protection to another order, and hence the localization length exponent $\nu$ can be larger than $2$ in this case.

More generally, we expect the exponent $\nu=2$ to hold whenever we have a family of drives that can be expanded in the form of Eq. \eqref{eq:generic expansion of U(t) near midpoint}.  In particular, let $\tilde{U}$ be any local unitary operator (not drive) that takes the block-diagonal form of \eqref{eq:Umidpoint as UAA+UBB} in some basis.  Suppose that the flow indices of the diagonal components satisfy $\nu(U_{AA})=-\nu(U_{BB})=1$.  Consider the family of time evolutions defined by
\beq
    \tilde{U}(\dt) = T\exp\left[ -i \int_0^{\dt} dt'\ H(t') \right] U_0,
\eeq
where $\tilde{H}(t)$ is some local Hamiltonian with no symmetries assumed.  Then we have the following expansion for small $\dt$:
\beq
    \tilde{U}(\dt) = \left[ 1 - i H(\dt=0)\dt +O((\dt)^2) \right]\tilde{U}_0.
\eeq
We claim that the localization length $\lloc$ of $\tilde{U}(\dt)$ follows the same exponent \eqref{eq:main claim} provided that the condition analogous to \eqref{eq:Hplus non-zero} holds, i.e.,
\beq
    \{ H(\dt=0), \tau^z \} \ne 0,\label{eq:H(0) not anticommute}
\eeq
where $\tau^z$ is the Pauli matrix acting on the space in which $\tilde{U}_0$ is block diagonal.

\section{Example drives with the universal exponent}\label{sec:Example drives with the universal exponent}
We now provide several example class AIII drives in which we find that the delocalization exponent is 2.  The starting point in these examples is a clean drive previously introduced in Ref. \cite{LiuChiral2018}.  With on-site disorder included, we demonstrate rigorously that the exponent is $\nu=2$ by using a result from the mathematics literature on products of random matrices \cite{SchraderPerturbative2004}.  In the special case of full phase disorder, we obtain an analytical result for the localization length at any time $t$ in the drive, which agrees with results found in similar models that have been studied in the context of discrete-time quantum walks.  We then allow disorder that couples neighboring sites within dimers, and we find that our prediction of $\nu=2$ is consistent with transfer matrix numerics.  Finally, we introduce disorder with exponential decay in position space, and we find that our prediction of $\nu=2$ is consistent with exact diagonalization numerics.  For further numerical checks, see Ref. \cite{BrownGeometric2019}.

We begin by recalling the drive introduced in Ref. \cite{LiuChiral2018}.  The model is a Floquet version of the Su-Schrieffer-Heeger (SSH) model \cite{SuSolitons1979}, and will be referred to hereafter as the SSH-type drive.  The drive occurs on an infinite one-dimensional bipartitate lattice with sublattice degrees of freedom labeled A and B.  The drive consists of piecewise-constant evolution by two Hamiltonians: $H_1$, which is an SSH Hamiltonian in the trivial phase, and $H_2$, which is an SSH Hamiltonian in the topological phase.  In particular, we define
\bseq
\begin{align} 
    H_1 &= \frac{2 \pi}{\td}\sum_n (\ket{n,A}\bra{n,B} + \text{H.c.}),\label{eq:H1} \\
    H_2 &= - \frac{2 \pi}{\td}\sum_n (\ket{n+1,A}\bra{n,B} + \text{H.c.}),\label{eq:H2}
\end{align}
\eseq
where $n$ is the site index and $\td$ is the period of the drive.  The time-dependent Hamiltonian is given by
\beq
  H(t) =
    \begin{cases}
      H_1 &  0\le t < \frac{\td}{4}\\
      H_2 &  \frac{\td}{4} \le t\le   \frac{3\td}{4}\\
      H_1 &  \frac{3\td}{4}<t\le  \td.
  \end{cases} \label{eq:model_drive}
\eeq
The time evolution operator for the drive is then given by the usual expression $U_\text{clean}(t) = T \exp[-i\int_0^t dt'\ H(t')]$.  As discussed in Ref. \cite{LiuChiral2018}, the one-dimensional AIII topological index (the chiral flow) for this drive is equal to $1$. This can be seen by noting that the time evolution operator at $t=\td/2$ acts separately on the $A$ and $B$ sublattices, and is given by
\begin{align}
     U_\text{clean}(t = \td/2) &= \begin{pmatrix} \transl_A & 0 \\ 0 & \transl^\dagger_B\end{pmatrix},
\end{align}
where $\transl_\alpha$ ($\alpha=A,B$) translates the corresponding sublattice by one unit cell to the right. Thus, the effect of the time evolution operator at the midpoint of the drive is the translation of all the $A$ orbital amplitudes by one unit cell to the right, and the translation of all the $B$ orbital amplitudes by one unit cell to the left.

The quasienergy spectrum of $U_\text{clean}(t)$ is gapless at $t=\td/2$ and gapped elsewhere (Fig. \ref{fig:spectrum}).  We present the explicit calculation in Appendix \ref{sec:Solution of the clean model} for completeness.  Alternatively, we can read off the solution by noting that the eigenstate problem for $U(t)$ [Eq. \eqref{eq:eigenstate eq}] is equivalent to a type of discrete-time quantum walk considered in Ref. \cite{VakulchykAnderson2017}.  Our parameter $\dt$ corresponds to the coin parameter of the quantum walk (see Appendix \ref{sec:Mapping to discrete-time quantum walk} for details).  
\begin{figure}[htb]
\centering
    \includegraphics[width=\columnwidth]{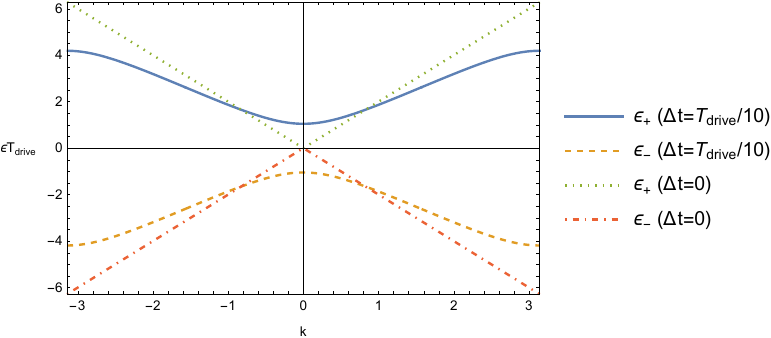}
    \caption{The two quasienergy bands $\epsilon_\pm(k)$ of the clean SSH-type drive.  For any $|\dt | < \td/4$, there is a gap, except at the midpoint of the drive ($\dt=0$, $\epsilon_{\pm}(k)= \pm 2|k| / \td$).  Here we have included the dimensionful factor $\td$ that we usually set to $2\pi$.}\label{fig:spectrum}
\end{figure}

We note here that if the model is considered with open boundary conditions [by restricting the sums over $n$ to $1,\dots, N$ in Eq. \eqref{eq:H1} and $1,\dots, N-1$ in Eq. \eqref{eq:H2}], then the time evolution operator for a full period acts as the identity in the bulk but not at the edges.  Indeed, sites at the edges ($B$ at $n=1$ and $A$ at $n=N$) gain a phase of $\pi$ under evolution by a full period \cite{LiuChiral2018}.  This micromotion at the edges is similar to the anomalous edge transport that occurs in the Rudner-Lindner-Berg-Levin model in two spatial dimensions \cite{RudnerAnomalous2013}.

In accordance with the general arguments of Sec. \ref{sec:Universal localization-delocalization transition}, we find that for a variety of disorderings of this drive, the localization length follows the behavior \eqref{eq:main claim}.  Let us note here that the localization length can be defined in various ways depending on the boundary conditions imposed on the model; the different definitions can be expected to become equivalent in the thermodynamic limit.  We present explicit definitions below.

Throughout the calculations below, we set $\td =2\pi$ for convenience.  We write an eigenstate $\ket{\Psi_\epsilon(t)}$ of $U(t)$ [c.f. Eq. \eqref{eq:eigenstate eq}] as
\beq
    \ket{\Psi_\epsilon(t)}= \sum_n (\Psi_{n,A}\ket{n,A} +\Psi_{n,B}\ket{n,B}),\label{eq:eigenstate parametrization}
\eeq
suppressing the $\epsilon$ and $t$ dependence of $\Psi_{n,A}$ and $\Psi_{n,B}$, and we define
\beq
    \Psi_n = 
    \bpmat
        \Psi_{n,A}\\
        \Psi_{n,B}
    \epmat.\label{eq:Psi spinor}
\eeq

\subsection{On-site disorder}\label{sec:On-site disorder}
\subsubsection{Setup}\label{sec:Setup}
We introduce two types of disorder into the SSH-type drive: phase disorder and bond disorder.  In the case of phase disorder only, we construct a disordered loop drive $U(t)$ using the approach described above Eq. \eqref{eq:disordered unitary one-sided}.  As we see below, phase disorder amounts to including a disordered Hamiltonian $H_d$ with random on-site energies.  In the case of bond disorder (possibly allowing both bond and phase disorder), we follow the approach described by Eq. \eqref{eq:generic expansion of U(t) near midpoint} and the discussion below there, in which we do not explicitly construct the full loop drive, but instead expand about the midpoint.  Bond disorder may loosely be thought of as disorder in the hopping amplitudes of the topologically nontrivial Hamiltonian $H_2$ in Eq. \eqref{eq:H2}.

To introduce phase disorder, we construct a disordered unitary operator $U_d$ and consider the drive $U(t)$ defined by Eq. \eqref{eq:disordered unitary one-sided}.  The disordered unitary is
\beq
     U_d = \sum_n \left( e^{i \phi_{n,A} } \ket{n,A}\bra{n,A} +e^{i \phi_{n,B} } \ket{n,B}\bra{n,B}\right),\label{eq:Ud phase disorder}
\eeq
where $\phi_{n,A}$ and $\phi_{n,B}$ are disordered phases.  In particular, we will assume that $\phi_{n,A}$ and $\phi_{n,B}$ are each independently and identically distributed (i.i.d.), with disorder distributions (one for the $A$ phases and another for the $B$ phases) that we need not specify at this point.

Let us verify that $U_d$ satisfies the required properties.  To do this, we must find a Hamiltonian $H_d$ that commutes with the chiral operator $\mathcal{C}$ and that yields $U_d$ through Eq. \eqref{eq:Ud general}.  We take $H_d$ to consist of disordered on-site energies:
\beq
    H_d = \frac{2}{T_0} \sum_n ( \phi_{n,A} \ket{n,A}\bra{n,A} + \phi_{n,B}\ket{n,B}\bra{n,B} ).
\eeq
Then it is clear that $H_d$ commutes with $\mathcal{C}$ and that Eq. \eqref{eq:Ud general} holds.

It is straightforward to show that the eigenstate equation \eqref{eq:eigenstate eq} is equivalent to \cite{BrownGeometric2019}
\bseq
\begin{align}
    e^{-i \epsilon t} \Psi_{n,A}&= \cos\left(\dt \right)e^{i \phi_{n-1,A}} \Psi_{n-1,A} \notag\\
    &\qquad \qquad + i \sin( \dt)e^{i \phi_{n,B}}  \Psi_{n,B}, \label{eq:SchrodA phase disorder}\\
    e^{-i \epsilon t} \Psi_{n,B}&= i \sin (\dt)e^{i \phi_{n,A}} \Psi_{n,A} \notag\\
     &\qquad \qquad+ \cos(\dt)e^{i \phi_{n+1,B}} \Psi_{n+1,B}.\label{eq:SchrodB phase disorder}
\end{align}
\eseq
This completes the setup of the SSH-type drive with phase disorder.  In this case, we have explicitly constructed a disordered, chiral-symmetric loop drive that can be studied at any $t$ in the driving period, though our main focus is on the regime of $t \approx \td/2$.

To include bond disorder, we do not explicitly construct the full loop drive for all $t$, but instead expand about the midpoint $t=\td/2$ [see Eq. \eqref{eq:generic expansion of U(t) near midpoint}].  We take $U(\td/2)$ and $H(\td/2)$ as given disordered operators.  We consider $U(\td/2)$ to be as in the clean SSH-type drive with phase disorder and $H(\td/2)$ to be $H_2$ from the SSH-type drive with disorder in the hopping amplitudes; that is,
\begin{widetext}
\bseq
\begin{align}
    U(\td/2) &=
    \sum_n ( \ket{n+1,A}\bra{n,A}  + \ket{n-1, B}\bra{n,B} )U_d,\\
    H(\td/2) &= - \sum_n v_n \left( \ket{n+1,A}\bra{n,B} + \text{H.c.} \right),
\end{align}
\eseq
\end{widetext}
where $U_d$ is given by Eq. \eqref{eq:Ud phase disorder} and where the $v_n$ variables are i.i.d. with some distribution.

By a straightforward calculation, we find that the eigenstate equation \eqref{eq:eigenstate eq} is equivalent to 
\bseq
\begin{align}
         e^{-i\epsilon t} \Psi_{n,A}&= e^{i\phi_{n-1,A}}\Psi_{n-1,A} \notag\\
         &\  + i e^{i\phi_{n,B}}v_{n-1} \dt \Psi_{n,B} + O((\dt)^2),\label{eq:SchrodA bond disorder} \\
        e^{-i\epsilon t} \Psi_{n,B}&= i e^{i\phi_{n,A}} v_n \dt \Psi_{n,A} \notag\\
        &\qquad + e^{i\phi_{n+1,B}}\Psi_{n+1,B}+O((\dt)^2).\label{eq:SchrodB bond disorder} 
\end{align}
\eseq
This completes the setup of the combination of bond and phase disorder in the approach of expanding about the midpoint.

We next consider a more general problem that can be specialized to either of the two setups above.  In particular, we consider the state determined by
\bseq
\begin{align}
        e^{-i\epsilon t} \Psi_{n,A}&= e^{i\phi_{n-1,A}}\cos(v_{n-1}\dt)\Psi_{n-1,A} \notag\\
        &\qquad + i e^{i\phi_{n,B}} \sin(v_{n-1}\dt) \Psi_{n,B},\label{eq:SchrodA extended} \\
        e^{-i\epsilon t} \Psi_{n,B}&= i e^{i\phi_{n,A}} \sin(v_n \dt) \Psi_{n,A} \notag\\
        &\qquad + e^{i\phi_{n+1,B}}\cos(v_n\dt)\Psi_{n+1,B}.\label{eq:SchrodB extended} 
\end{align}
\eseq
By studying Eqs. \eqref{eq:SchrodA extended} and \eqref{eq:SchrodB extended}, we can simultaneously consider both the case of phase disorder (on a disordered loop drive defined for all $\dt$) and the case of a combination of phase and bond disorder (defined explicitly only for small $\dt$).  To see this, we note that setting all $v_n=1$ recovers Eqs. \eqref{eq:SchrodA phase disorder} and \eqref{eq:SchrodB phase disorder}, while expanding in $\dt$ recovers Eqs. \eqref{eq:SchrodA bond disorder} and \eqref{eq:SchrodB bond disorder}.

Equations \eqref{eq:SchrodA extended} and \eqref{eq:SchrodB extended} may be brought to the following transfer matrix form:
\beq
    \Psi_{n+1}
    = \mathcal{M}_n
    \Psi_n,\label{eq:position-space transfer matrix def}
\eeq
where the transfer matrix is
\begin{widetext}
\beq
    \mathcal{M}_n = \bpmat
        e^{i (\epsilon t+\phi_{n,A})}\sec (v_n \dt)  & i \tan(v_n \dt) \\
         -i e^{i(\phi_{n,A}-\phi_{n+1,B})}\tan (v_n \dt)  & e^{-i( \epsilon t +\phi_{n+1,B})} \sec (v_n \dt)
    \epmat.\label{eq:transfer matrix phase and bond disorder}
\eeq
\end{widetext}

Using the mapping to discrete-time quantum walks that we present in Appendix \ref{sec:Mapping to discrete-time quantum walk}, we can show that the transfer matrix \eqref{eq:transfer matrix phase and bond disorder} is a special case of a transfer matrix obtained in Ref. \cite{VakulchykAnderson2017}.  Note that, although a $4\times 4$ transfer matrix might have been expected (since we have a bipartite lattice with nearest-neighbor coupling), there is in fact a $2\times 2$ transfer matrix \cite{VakulchykAnderson2017}.  For the case of phase disorder only, Eq. \eqref{eq:transfer matrix phase and bond disorder} was also obtained by some of the present authors in Ref. \cite{BrownGeometric2019}.

\subsubsection{Localization length---calculation in position space}\label{sec:Localization length---calculation in position space}
For our first calculation of the localization length for the problem specified by Eqs. \eqref{eq:SchrodA extended} and \eqref{eq:SchrodB extended}, we use a position space approach.  The time-dependent Lyapunov exponent $\gamma(t)$ is defined by
\beq
    \gamma(t) = \lim_{N\to\infty} \frac{1}{N}\langle \ln || \mathcal{M}_N\dots \mathcal{M}_1 || \rangle_{1\dots N},\label{eq:Lyapunov exponent definition}
\eeq
where the double brackets indicate a matrix norm (e.g., the $2$-norm) and the subscripts $1,\dots, N$ indicate the average over all disorder variables associated with those sites ($\phi_{1,A},\dots,\phi_{N,A}$, etc.).  The localization length and Lyapunov exponent are related by $1/\lloc(t)=\gamma(t)$.

Products of random matrices have been extensively studied in the mathematics literature.  In particular, a result by Schrader \emph{et al}. in Ref. \cite{SchraderPerturbative2004} can be applied to our present problem to yield the Lyapunov exponent to the leading order in $\dt$.  The final result is (see Appendix \ref{sec:Position space calculation} for details)
\begin{widetext}
\beq
    \frac{1}{\lloc(t)}= \frac{1}{2}\biggr(\langle v_n^2 \rangle_n + 2\text{Re} \biggr[ \frac{\langle v_n \rangle_n^2\langle e^{i\phi_{n,A} }\rangle_n  \langle e^{i\phi_{n,B} }\rangle_n}{e^{-2\pi i \epsilon} -\langle e^{i\phi_{n,A} }\rangle_n  \langle e^{i\phi_{n,B} }\rangle_n }    \biggr] \biggr)(\dt)^2 + O((\dt)^3),\label{eq:lloc model drive phase and bond disorder}
\eeq
\end{widetext}
where $n$ is any site and $\langle \dots \rangle_n$ indicates the average over the disorder variables of that site ($\phi_{n,A}$, $\phi_{n,B}$, and $v_n$).  Note that the leading term is proportional to $(\Delta t)^2$.  This demonstrates that for the SSH-type drive with the on-site disorder that we have considered in this section, our main claim [Eq. \eqref{eq:main claim}] holds.

\subsection{Phase disorder in dimers}\label{sec:Phase disorder in dimers}
As a further probe of the universality of the localization length exponent, we introduce a type of phase disorder into the SSH-type drive that extends beyond individual sites.   We group adjacent sites $(2n-1,2n)$ (for any integer $n$) into dimers and introduce phase disorder that couples the two sites within each dimer.  We find that this drive may be described by a $2\times 2$ transfer matrix.  We then study this transfer matrix numerically and show that the localization length has an exponent consistent with $\nu=2$.

We introduce dimer phase disorder as a natural generalization of the on-site phase disorder considered in Sec. \ref{sec:On-site disorder}.  In particular, we allow the disordered Hamiltonian $H_d$ to include both on-site terms and terms that couple sites within the same dimer (i.e., the sites $2n-1$ and $2n$).  For $H_d$ to commute with the chiral symmetry operator, as is required for the construction of Sec. \ref{sec:Chiral-symmetric disorder in loop drives}, the $A$ amplitudes must not be coupled with the $B$ amplitudes.  The most general such $H_d$ acts as some Hermitian matrix on each pair $\ket{2n-1,A}$ and $\ket{2n,A}$, and as another Hermitian matrix on each pair $\ket{2n-1,B}$ and $\ket{2n,B}$.  The disordered unitary operator $U_d = e^{-i H_d T_0/2}$ then acts as some unitary matrix $U_{n,A}$ on each pair $\ket{2n-1,A}$ and $\ket{2n,A}$, and as another unitary matrix $U_{n,B}$ on each pair $\ket{2n-1,B}$ and $\ket{2n,B}$.

We find it more convenient to formulate the problem in terms of the matrix elements of $U_{n,A}$ and $U_{n,B}$, rather than in terms of the parameters of $H_d$ that generate these matrices.  Parametrizing the disordered unitary matrices as
\beq
    U_{n,\alpha} = 
    \bpmat
        U_{n,\alpha}^{(1,1)} & U_{n,\alpha}^{(1,2)}\\
        U_{n,\alpha}^{(2,1)} & U_{n,\alpha}^{(2,2)}
    \epmat
    \qquad (\alpha = A \text{ or }B),
\eeq
we then write disordered unitary operators for each dimer as
\beq
    \mathcal{U}_{n,\alpha}=\sum_{j,m=1,2} \ket{2n-2+j,\alpha}\bra{2n-2+m,\alpha} U_{n,\alpha}^{(j,m)},
\eeq
where $\alpha=A$ or $B$.  The disordered unitary is then a sum over the dimers:
\beq
    U_d = \sum_n ( \mathcal{U}_{n,A} + \mathcal{U}_{n,B}).
\eeq
We do not yet specify the disorder distribution but will assume that $U_{n,A}$ and $U_{n,B}$ separately are i.i.d. across the dimers labeled by $n$.

Then, Eq. \eqref{eq:disordered unitary one-sided} defines a chiral-symmetric loop drive $U(t)$ with phase disorder within each dimer, generalizing the on-site case considered in the previous section.  We now set up the transfer matrix to do numerical calculations of the localization length.  As the calculation is lengthy, we defer details to the Appendix and summarize the result as follows.  The eigenstate equation for the drive [Eq. \eqref{eq:eigenstate eq}] may be written equivalently as a transfer matrix equation involving two of the four amplitudes in adjacent dimers, and a second equation that relates the amplitudes within one dimer:
\bseq
\begin{align}
    \bpmat
        \Psi_{2n+1,A}\\
        \Psi_{2n+2,B}
    \epmat
    = 
        \mathcal{M}_{n,\text{dimer}}
    \bpmat
        \Psi_{2n-1,A}\\
        \Psi_{2n,B}
    \epmat,\label{eq:transfer matrix relation dimer disorder}\\
    \bpmat
        \Psi_{2n-1,B}\\
        \Psi_{2n,A}
    \epmat
    = \mathcal{M}_{n,\text{dimer}}'
    \bpmat
        \Psi_{2n-1,A}\\
        \Psi_{2n,B}
    \epmat.\label{eq:Mprime}
\end{align}    
\eseq
The transfer matrix  $\mathcal{M}_{n,\text{dimer}}$ is a function of $U_{n,A}$, $U_{n+1,A}$, $U_{n,B}$, $U_{n+1,B}$, $\epsilon$, and $t$ (see Appendix \ref{sec:Transfer matrix for the SSH-type drive with dimer disorder} for the explicit expression), and the Lyapunov exponent $\gamma(t)$ is given by Eq. \eqref{eq:Lyapunov exponent definition} with $\mathcal{M}_n$ replaced by $\mathcal{M}_{n,\text{dimer}}$.  The explicit form of the other matrix ($\mathcal{M}_{n,\text{dimer}}'$) is not needed.

\begin{figure}
    \centering
    \includegraphics[width=0.9\columnwidth]{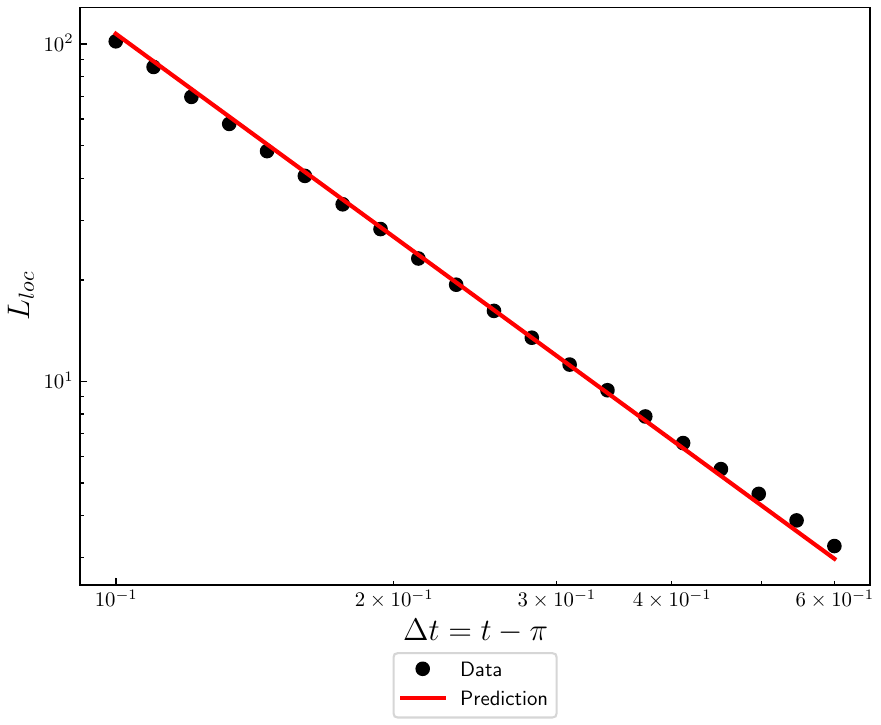}
    \caption{The localization length in the SSH-type drive with the dimer disorder described in Sec.~\ref{sec:Phase disorder in dimers}. The disordered matrices $\mathcal{U}_{n,A}$ and $\mathcal{U}_{n,B}$ were chosen uniformly from $U(2)$ using the Haar measure. The localization length at quasienergy $\epsilon$ is calculated according to $\lloc(t)= \gamma(t)^{-1}$, wherein $\gamma(t)$ is the Lyapunov exponent defined in Eq.~\eqref{eq:Lyapunov exponent definition}, with the matrix norm chosen to be the spectral norm. The localization length at each value of $\dt$ is computed for $\epsilon=0.1$ using $L=10^{6}$ transfer matrices. The prediction curve is a linear fit assuming $\nu=2$ [i.e.,  $1/\lloc(t) = c (\dt)^2$ with fit parameter $c$].}
    \label{fig:dimer_disorder_plot1}
\end{figure}

We have thus brought the problem to a transfer matrix form, which is convenient for numerics.  We calculate the Lyapunov exponent numerically at several values of $\dt$, with results consistent with our prediction of $\nu=2$ (see Fig. \ref{fig:dimer_disorder_plot1}).
It does not seem feasible, though, to obtain an analytical answer using the result from Ref. \cite{SchraderPerturbative2004}, because this result requires the disorder to separate across the transfer matrices, whereas here the disordered matrices $U_{n,A}$ and $U_{n,B}$ appear in both $\mathcal{M}_{n,\text{dimer}}$ and $\mathcal{M}_{n-1,\text{dimer}}$.

\subsection{Longer-range disorder}
As a further test of the universality of the exponent $\nu=2$, we do a numerical test using a version of disorder that extends beyond dimers.  In particular, we consider disorder that decays in exponentially in real space.  Due to this longer range, we cannot set up a transfer matrix calculation.  We therefore impose periodic boundary conditions and use exact diagonalization to calculate the localization length numerically (see below).

We again define a disordered drive by Eqs. \eqref{eq:Ud general} and \eqref{eq:disordered unitary one-sided}, this time using a disordered Hamiltonian $H_d$ that couples nearest neighbors.  In particular, we take $H_d$ to be an independent copy of the static Anderson model (nearest-neighbor hopping with disordered on-site energies) on each sublattice:
\begin{multline}
    H_d = \sum_{n,\alpha} \epsilon_{n\alpha}\ket{n,\alpha}\bra{n,\alpha}\\
    - V \sum_{n,\alpha} \left( \ket{n+1,\alpha} \bra{n,\alpha} + \text{H.c.}\right).\label{eq:Hd Anderson}
\end{multline}
Since the two sublattices are decoupled, we have the required condition that $H_d$ commutes with the chiral symmetry.  Although $H_d$ only couples nearest neighbors, $U_d$ extends farther because it is obtained by exponentiating $H_d$.  (In fact, it can be shown that $U_d$ must be exponentially local in real space \cite{GrafBulk2018}.)

\begin{figure}[htb]
\centering
    \includegraphics[width=0.9\columnwidth]{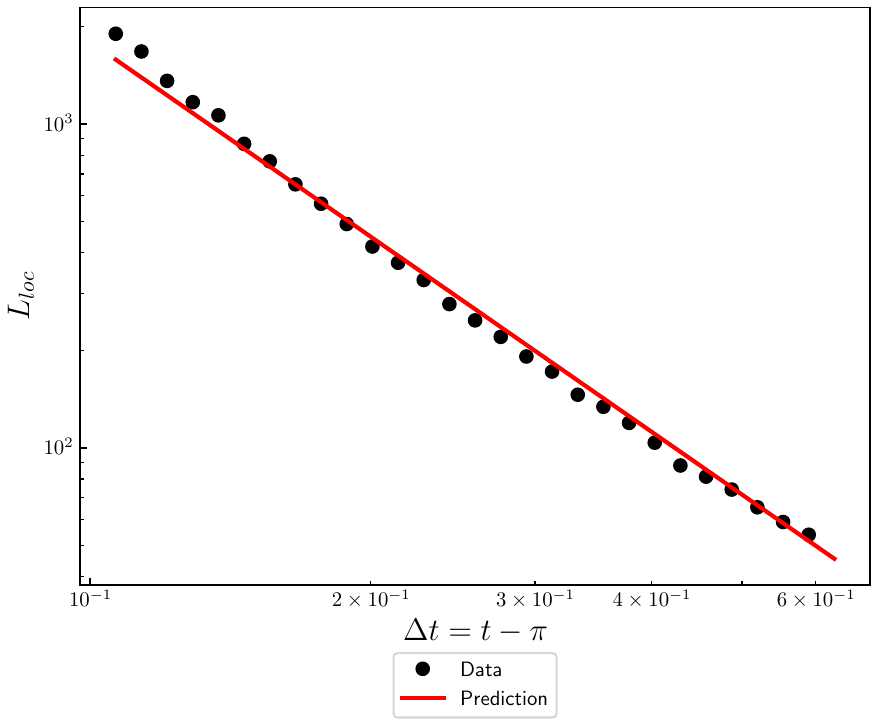}
    \caption{Test of the universal exponent in the SSH-type drive with the longer-range disorder obtained from Eq. \eqref{eq:Hd Anderson}.  The prediction curve is a linear fit assuming $\nu=2$ [i.e.,  $1/\lloc(t) = c (\dt)^2$ with fit parameter $c$].}\label{fig:Anderson_plot}
\end{figure}

From Eq. \eqref{eq:Ud general}, we see that the dimensionless quantities that appear in $U_d$ are $\epsilon_{n\alpha}T_0$ and $V T_0$.  We consider uniform on-site disorder; in particular, we take $\epsilon_{n\alpha}T_0/2$ to be uniformly distributed (independently on each site and sublattice) in $[-\pi,\pi]$.  We also set $VT_0/2=1$.  To reduce numerical noise, we average the localization length over all quasienergies.  Due to this averaging, we are testing a weaker version of \eqref{eq:main claim}.  The numerical results are consistent with the exponent $\nu=2$ (Fig. \ref{fig:Anderson_plot}).

The numerical calculation of the localization length is done as follows.  At a given value of $t$, we consider a sequence of system sizes $L$.  At each $L$, we use exact diagonalization to find all eigenstates of $U(t)$.  The $L$-dependent localization length $\lloc(t,L)$ of each eigenstate $\ket{\Psi_\epsilon(t)}$ is defined as the root-mean-square variation in position space:
\beq
    \lloc(t,L) = \sqrt{\bra{\Psi_\epsilon(t)} (\hat{X}- \overline{X})^2 \ket{\Psi_\epsilon(t)}},
\eeq
where $\hat{X}$ is the position operator:
\beq
    \hat{X} = \sum_{n=1}^L\sum_{\alpha=A,B} n \ket{n,\alpha}\bra{n,\alpha},
\eeq
and where the mean value $\overline{X}$ is defined in the appropriate way for periodic boundary conditions \cite{RestaQuantumMechanical1998}:
\beq
    \overline{X}= \frac{L}{2\pi} \text{Im}\left[ \ln\bra{\Psi_\epsilon(t)} e^{2\pi i \hat{X}/L} \ket{\Psi_\epsilon(t)} \right]
    .
\eeq

We are interested in the localization length in the thermodynamic limit, that is, $\lloc(t) \equiv \lim_{L\to\infty} \lloc(t,L)$. 
To calculate this, we fit the $L$-dependent data to $\lloc(t,L)=\lloc(t) - a e^{-b L}$ with fit parameters $\lloc(t)$, $a$, and $b$.  (For smaller values of $L$, we also average over multiple disorder realizations.)  This procedure is done for each eigenstate, and the average of $\lloc(t,\epsilon)$ over all eigenstates yields a data point in Fig. \ref{fig:Anderson_plot}.

\section{Scattering argument for the universal exponent}\label{sec:Scattering argument for the universal exponent}

In the thermodynamic limit, it can be expected that the localization length should not depend on the choice of boundary conditions.  We find that scattering boundary conditions (defined below) are convenient for making an analytical argument for the universality of the exponent $\nu=2$.

To set up the scattering problem, we write the time-dependent Hamiltonians that generate the drives $U(t)$ and $U_\text{clean}(t)$ as $H(t)$ and $H_\text{clean}(t)$, respectively.  We assume that the disorder in $H(t)$ is short-range correlated, in the following sense: We have
\beq
    H(t) = H_\text{clean}(t)+ H_\text{disorder}(t),\label{eq:H_for_scattering}
\eeq
where
\beq
    H_\text{disorder}(t) = \sum_n \Theta_n(t),\label{eq:Hdisorder}
\eeq
and each operator $\Theta_n(t)$ is exponentially localized near the site $n$.    Furthermore, the decay lengths are upper bounded by a number that is constant in the thermodynamic limit and independent of $t$.

We then define a disordered ``sample,'' consisting of the sites $n=1,\dots, N$, by setting $\Theta_n(t)= 0$ for all $n \le 0$ and $n \ge N+1$.  The $\Theta_n(t)$ operators in the sample are i.i.d. with some probability distribution.  The infinite regions to the right and left of the sample are the ``leads.''  Although the matrix elements of the corresponding unitaries, $U(t)$ and $U_\text{clean}(t)$, need not be exactly equal, they can be expected to be asympotically equal: $U_{nn'}(t) \to U_{\text{clean},nn'}(t)$ far from the disordered region (with error exponentially suppressed in the distance).  This permits a scattering treatment for the eigenstate wavefunctions of $U(t)$.

Let us pause here for a technical point.  It would be more natural to take the disorder to be of the form of Eq. \eqref{eq:Hdisorder} for the Hamiltonian $\mathcal{H}(t)$ that generates the original drive $\mathcal{U}(t)$, rather than for the Hamiltonian $H(t)$ that generates the loop drive $U(t)$.  Given the complexity of the loop decomposition, it is not clear that Eq. \eqref{eq:Hdisorder} would hold in this case.  However, although we assume Eq. \eqref{eq:Hdisorder} here to make the scattering setup clearer, it is a stronger assumption than necessary; for our arguments below, it suffices to assume that $U(t)$ is amenable to a scattering treatment (with any number of consecutive sites being chosen to define a disordered sample).  We have verified this assumption numerically in a simple example of the more natural disorder setup (see Appendix \ref{sec:Disorder remains short-ranged under loop decomposition}).

We fix $t$ and consider the scattering problem for $U(t)$.  Incoming and outgoing waves in the leads are described by scattering amplitudes (complex numbers), and these amplitudes are related by the S matrix of the sample (a unitary matrix denoted $S_{1\dots N}$).  We focus on the case that the S matrix is $2\times 2$.  Then, there are four scattering amplitudes that may be labeled as $\Psi_\alpha^\pm$ ($\alpha=L,R$), where $+$ ($-$) refers to right-moving (left-moving) waves and $L$ and $R$ refer to the left and right leads.  The scattering amplitudes are related by
\beq
    \bpmat
        \Psi_R^+ \\
        \Psi_L^-
    \epmat
    = \mathcal{S}_{1\dots N}
    \bpmat
        \Psi_L^+\\
        \Psi_R^-
    \epmat.\label{eq:S matrix general}
\eeq 
Below, we argue that the $2\times 2$ case suffices to cover a wide class of drives, and we also provide a more explicit definition of the four scattering amplitudes. 

The S matrix may be parametrized by transmission and reflection amplitudes:
\beq
    \mathcal{S}_{1\dots N} =
    \bpmat
        t_{1\dots N} & r_{1\dots N}'\\
        r_{1\dots N} & t_{1\dots N}'
    \epmat.\label{eq:S matrix general parametrization}
\eeq

In the scattering setup, localization manifests as the exponential decay of the typical transmission coefficient $T_{1\dots N}\equiv |t_{1\dots N}|^2 = |t_{1\dots N}'|^2$ as $N$ increases: $T_{1\dots N}^{\text{(typical)}} \sim e^{-2N/\lloc}$.  From the theory of disordered systems in one dimension, the distribution of $T_{1\dots N}$ over disorder realizations becomes log-normal for large $N$, and hence it is the average of the logarithm of the transmission coefficient that determines the typical value.  In particular, we have
\beq
     \frac{2}{\lloc}= \lim_{N\to\infty} \frac{1}{N}\langle - \ln T_{1\dots N} \rangle_{1\dots N}.
\eeq

Unless $T_{1\dots N}= 0$ (which we assume can only occur on a set of measure zero among disorder realizations), the basic relation \eqref{eq:S matrix general} can also be written in terms of the scattering transfer matrix $\mathcal{T}_{1\dots N}$ as
\beq
    \bpmat
        \Psi_R^+\\
        \Psi_R^-
    \epmat
    =
    \mathcal{T}_{1\dots N}
    \bpmat
        \Psi_L^+\\
        \Psi_L^-
    \epmat.\label{eq:scattering transfer matrix general}
\eeq
The unitarity of the S matrix is equivalent to the following pseudounitarity condition:
\beq
    \mathcal{T}_{1\dots N}^\dagger \sigma^z \mathcal{T}_{1\dots N} = \sigma^z.\label{eq:pseudo-unitarity condition}
\eeq
For later use, let us recall here the general parametrization of a scattering transfer matrix:
\beq 
    \mathcal{T}_{1\dots N} = 
    \bpmat
        1/t_{1\dots N}^* & r_{1\dots N}' / t_{1\dots N}' \\
        -r_{1\dots N}/t_{1\dots N}' & 1/t_{1\dots N}'
    \epmat.\label{eq:scattering transfer matrix general parametrization}
\eeq

The basic tool in our arguments below is the following formula from Refs. \cite{CulverScattering2024a} and\cite{CulverScattering2024}, which (as discussed there) is a corollary to the result from Schrader \emph{et al}. \cite{SchraderPerturbative2004} that we used in Sec. \ref{sec:Localization length---calculation in position space}.  The formula applies to the case of scattering that factorizes into a product of individual scattering events with i.i.d. disorder.  In particular, suppose that the scattering transfer matrix of the sample can be written in the product form
\beq
    \mathcal{T}_{1\dots N} = \mathcal{T}_{N_s}\dots\mathcal{T}_1,\label{eq:scattering transfer matrix factorization}
\eeq
where each $\mathcal{T}_j$ is a $2\times 2$ scattering transfer matrix [i.e., it satisfies Eq. \eqref{eq:pseudo-unitarity condition}] and where $N_s$ is the number of scattering events.  (In the simplest case, $N_s=N$, but we later consider the more general case of $N_s\le N$.)  Suppose further that the entries of $\mathcal{T}_j$ ($j=1,\dots, N_s$) are i.i.d.; note that this amounts to assuming that the disorder in $U(t)$ is uncorrelated in space and sufficiently short-ranged.  Let $\mathcal{T}_j$ be parametrized as in Eq. \eqref{eq:scattering transfer matrix general parametrization} (in particular, with reflection amplitudes $r_j$ and $r_j'$ and reflection coefficient $R_j=|r_j|^2=|r_j'|^2$).  Then we have \footnote{This formula is based on assumptions that can be expected to hold for ``generic'' disorder distributions and model parameters.  The $O(|r_j|^3)$ can in fact be replaced by $O(|r_j|^4)$ (i.e., the third-order terms vanish), as discussed in Refs. \cite{CulverScattering2024a} and \cite{CulverScattering2024}; however, only the second-order terms are important for our present goal of showing $\nu=2$.}
\beq
    \frac{2}{\lloc} = \langle R_j\rangle_j - 2 \text{Re}\left[ \frac{\langle r_j \rangle_j \langle r_j'\rangle_j}{1+ \langle r_j r_j'/R_j\rangle_j } \right]+O(|r_j|^3),\label{eq:lloc to 2nd order}
\eeq
where the angle brackets indicate the disorder average over any $j=1,\dots, N_s$.  (We emphasize that the index $j$ generally stands for the disorder variables of a number of consecutive lattice sites.)

The key point is that both terms on the right-hand side are of second order in $|r_j|=|r_j'|$.  Thus, provided we can bring the scattering problem to the above form, it suffices to show $|r_j|\sim \dt$, for then Eq. \eqref{eq:lloc to 2nd order} yields the exponent $\nu=2$ in Eq. \eqref{eq:main claim}.

We separate the argument into two main parts.  First, we consider a particular model: the SSH-type drive with on-site disorder (the same model from Sec. \ref{sec:Example drives with the universal exponent}, now with scattering boundary conditions).  This model serves as a soluble starting point for later generalization.  We obtain the required factorization into scattering transfer matrices, and we thus recover the analytical expression \eqref{eq:lloc model drive phase and bond disorder} for the inverse localization length that we obtained above using a position-space approach.  In this simple problem, the number of scatterers is the same as the number of sites in the disordered sample: $N_s=N$.

Second, we argue that the exponent $\nu=2$ can be obtained in a wider class of drives.  Here we make the following modification (described in more detail below) to the scattering setup: We take the sample to consist of an alternating sequence of disordered regions and clean regions, with the disordered sites being a fixed fraction $f$ of the total number of sites in the sample.  Note that setting $f=1$ would recover the scattering problem as stated above.  In this setting, we obtain the exponent $\nu=2$ for any value $0< f <1$ by taking the scatterers in \eqref{eq:scattering transfer matrix factorization} to be the disordered regions within the sample (hence, $N_s<N$ in this case).  Equation \eqref{eq:lloc to 2nd order} becomes more difficult to evaluate explicitly, so we do not obtain any formula for the nonuniversal prefactor in Eq. \eqref{eq:main claim}; however, we still obtain $|r_j|\sim\dt$ and hence $\nu=2$.  We also state the assumption that would be needed to obtain $\nu=2$ in the limit $f\to1$.

\subsection{Scattering calculation for the SSH-type drive with on-site disorder}\label{sec:Scattering calculation for the SSH-type drive with on-site disorder}
We consider the clean drive, as defined in Sec. \ref{sec:Example drives with the universal exponent}, on an infinite lattice.  It is straightforward to modify the phase disorder and bond disorder constructions of Sec. \ref{sec:On-site disorder} to the scattering setup, as we now show.

For phase disorder, we set the phases $\phi_{n,A}=\phi_{n+1,B}=0$ except for $n$ in the sample ($n=1,\dots,N$).  Note that the $B$ phase is offset by one unit; this is done to simplify the transfer matrices at the sample edges, and there is no effect on the large $N$ behavior.  For bond disorder, we set $v_n=0$ except for $n=1,\dots,N$.  Both cases can be treated at once by using the transfer matrix defined by Eqs. \eqref{eq:position-space transfer matrix def} and \eqref{eq:transfer matrix phase and bond disorder}.

We fix a time $t$ within the drive, near but not equal to $\td/2$, and we consider an eigenstate $\ket{\Psi_\epsilon(t)}$ of $U(t)$ with quasienergy $\epsilon\ne 0$ [Eq. \eqref{eq:eigenstate eq}].  The clean spectrum becomes gapless as $\dt\to 0$; for sufficiently small $\dt$, the spectrum is doubly degenerate, with two momenta ($k$ and $-k$) corresponding to $\epsilon$ (see Fig. \ref{fig:spectrum}).  The eigenstates of $U_\text{clean}(t)$ are Bloch waves defined by [recalling Eqs. \eqref{eq:eigenstate parametrization} and \eqref{eq:Psi spinor}]
\beq
    \Psi_n = u(\pm k) e^{\pm i k n},
\eeq
where the Bloch functions $u(\pm k)$ are two-component spinors in the sublattice basis: $u(\pm k) = (u^A(\pm k),u^B(\pm k) )$.  There are two bands; for definiteness, we take $\epsilon>0$, so we are considering the upper band.  (Explicit expressions are given in Appendix \ref{sec:Solution of the clean model}.)

A scattering eigenstate can be written as a linear combination of Bloch waves in the leads.  The coefficients are the scattering amplitudes $\Psi_\alpha^\pm$ from Eq. \eqref{eq:S matrix general}.  Thus,
\begin{widetext}
\beq
    \Psi_n
    =
    \begin{cases}
        \Psi_L^+ u(k) e^{i k (n-1)} + \Psi_L^- u(-k) e^{-i k (n-1)} & n \le 1  \\
        \Psi_R^+ u(k) e^{i k (n-1-N)} + \Psi_R^- u(-k) e^{-i k (n-1-N)} & n \ge N+1.\label{eq:scattering wavefunction SSH-type drive}
    \end{cases}
\eeq
\end{widetext}
Our phase conventions here (i.e., the factors of $e^{\pm i k}$ and $e^{\pm i k N}$) are chosen for convenience.  

To obtain the factorization \eqref{eq:scattering transfer matrix factorization}, we define a matrix $\Lambda$ for converting from the basis of scattering amplitudes to the basis of position states:
\beq
    \Lambda =
    \bpmat
        u^A(k) & u^A(-k)\\
        u^B(k) & u^B(-k)
    \epmat.\label{eq:matLambda}
\eeq
We then have
\bseq
\begin{align}
    \Psi_1 &= \Lambda
    \bpmat
    \Psi_L^+\\
    \Psi_L^-
    \epmat,\\
    \Psi_{N+1} &= \Lambda
    \bpmat
    \Psi_R^+\\
    \Psi_R^-
    \epmat,
\end{align}
\eseq
and hence, by Eq. \eqref{eq:position-space transfer matrix def}, we obtain the scattering transfer matrix \eqref{eq:scattering transfer matrix general}:
\beq
    \mathcal{T}_{1\dots N}
    = \Lambda^{-1} \mathcal{M}_N\dots\mathcal{M}_1\Lambda.
\eeq
We thus obtain the factorized form \eqref{eq:scattering transfer matrix factorization} with $N_s=N$ and
\beq
    \mathcal{T}_n = \Lambda^{-1}\mathcal{M}_n \Lambda,\label{eq:matTn}
\eeq
where we have identified the scatterer index $j$ with the site index $n$ (because $N_s=N$).  It is readily checked that $\mathcal{T}_n$ satisfies the pseudo-unitary condition \eqref{eq:pseudo-unitarity condition}.

To apply Eq. \eqref{eq:lloc to 2nd order}, we parametrize $\mathcal{T}_n$ by $r_n,r_n'$, etc., as in Eq. \eqref{eq:scattering transfer matrix general parametrization}.  It is then straightforward to calculate $r_n$ and $r_n'$ and to then use Eq. \eqref{eq:lloc to 2nd order} to again obtain the leading order expression \eqref{eq:lloc model drive phase and bond disorder} (see Appendix \ref{sec:Scattering calculation} for details).

There is a simpler way to recover Eq. \eqref{eq:lloc model drive phase and bond disorder}: we note that $\mathcal{M}_n$ itself satisfies the pseudo-unitarity condition (i.e., we have $\mathcal{M}_n^\dagger \sigma^z\mathcal{M}_n = \sigma^z$).  We can thus consider an auxiliary problem which is defined by taking the scattering transfer matrix for the sample to be
\beq
    \widetilde{\mathcal{T}}_{1\dots N}= \widetilde{\mathcal{T}}_N\dots \widetilde{\mathcal{T}}_1,\label{eq:auxiliary problem SSH-type drive}
\eeq
where $\widetilde{\mathcal{T}}_n= \mathcal{M}_n$.  The localization length must be the same as the original problem because the scattering transfer matrices for the sample only differ by boundary terms (see, e.g., Appendix H of Ref. \cite{CulverScattering2024}): $\mathcal{T}_{1\dots N}= \Lambda^{-1} \widetilde{\mathcal{T}}_{1\dots N} \Lambda$.  We parametrize $\widetilde{\mathcal{T}}_n$ by $\tilde{r}_n,\tilde{r}_n'$, etc., as in Eq. \eqref{eq:scattering transfer matrix general parametrization}.  From Eq. \eqref{eq:transfer matrix phase and bond disorder}, we see that $\tilde{r}_n = -i e^{i(\epsilon t + \phi_{n,A})}\sin(v_n\dt)$ and $\tilde{r}_n'= i e^{i(\epsilon t + \phi_{n+1,B})}\sin(v_n\dt)$; then applying Eq. \eqref{eq:lloc to 2nd order} and expanding in $\dt$ indeed recovers Eq. \eqref{eq:lloc model drive phase and bond disorder}.

\begin{figure}
    \centering
    \includegraphics[width=0.8\columnwidth]{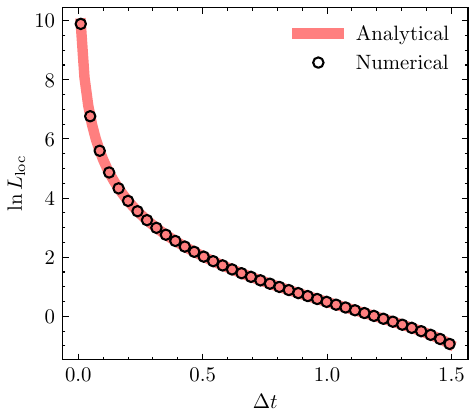}
    \caption{The localization length $\lloc(t)$ at quasienergy $\epsilon=0.57$ in the SSH-type drive with full phase disorder.  The numerical results match the analytical expression \eqref{eq:Lloc any dt model drive}. For each value of $\dt$, a single realization of disorder was used, and $\lloc(t)$ was calculated by obtaining the slope $s$ of a linear fit for $-\ln \widetilde{T}_{1\dots N}$ vs $N$ [in the auxiliary problem defined by Eq. \eqref{eq:auxiliary problem SSH-type drive}], and using the relation $s = \frac{2}{\lloc(t)}$. }\label{fig:logcosdt}
\end{figure}

As an aside, we consider the special case of full phase disorder, i.e., $\phi_{n,A}$ and $\phi_{n,B}$ are i.i.d. in $[-\pi,\pi)$ (and all $v_n=1$).  We can then calculate the localization length for arbitrary $t$ via the uniform phase formula \cite{AndersonNew1980}: $2/\lloc(t)= \langle - \ln \tilde{T}_n\rangle_n$, where $\tilde{T}_n$ is the transmission coefficient of the auxiliary problem (see Appendix \ref{sec:Full phase disorder} for details).  The result is
\beq
    1/\lloc(t) = |\ln \cos \dt|,\label{eq:Lloc any dt model drive}
\eeq
which agrees with numerics (Fig. \ref{fig:logcosdt}).
An equivalent formula to \eqref{eq:Lloc any dt model drive} has been obtained in related calculations for discrete-time quantum walks \cite{VakulchykAnderson2017,DerevyankoAnderson2018}.

\subsection{Generalization to other drives}\label{sec:Generalization to other drives}

The calculation of the previous section has some features that are not representative of the generic case.  First, the disordered unitary $U(t)$ there only couples nearest neighbors, while for a generic drive, $U(t)$ can extend further (generally it decays exponentially).  Second, the clean unitary $U_\text{clean}(t)$ of the SSH-type drive is non-generic both in its band structure and its lack of eigenstates with complex momenta (see below).

We argue that the universal exponent $\nu=2$ is still obtained even if both of these features are relaxed.  Without claiming to establish this result rigorously, we present a sequence of generalizations beyond the soluble model of the previous section, indicating the assumptions needed at each step.  Mainly, we need to modify the original scattering problem by making the disordered sample have some arbitrarily small fraction of clean regions, and we need to make assumptions about the band structure of $U_\text{clean}(t)$.

\paragraph{Beyond nearest neighbors.}
Let us first consider the generalization to disorder that connects sites beyond nearest neighbors.  Here, we continue to take $U_\text{clean}(t)$ to be the SSH-type drive as defined above.  We assume for now that the disordered drive $U(t)$ is strictly local; that is, there some range $\xi_\text{strict}$ for which
\beq
    U_{nn'}(t) = 0 \qquad (|n-n'| > \xi_\text{strict}),\label{eq:U(t) strictly local}
\eeq
where we suppress sublattice indices and any other on-site degrees of freedom.  (We later make a heuristic generalization to the case of exponential locality.)  Also, we assume that the constant $\xi_\text{strict}$ is independent of $t$.  The dimer disorder from Sec. \ref{sec:Phase disorder in dimers} provides an example of a drive with $\xi_\text{strict}=2$.

Due to the strict locality condition \eqref{eq:U(t) strictly local}, the disordered unitary agrees exactly with the clean unitary throughout the leads, except for regions of size $\xi_\text{strict}$ at the edges of the sample.  A scattering eigenstate can therefore be written in the leads as a linear combination of Bloch waves, provided that the edge regions are avoided.  In anticipation of later generalizations, let us define $k_\pm= \pm k$ and $\ell_\text{max}=\xi_\text{strict}$.  Then, a scattering eigenstate of $U(t)$ may be written in the leads as
\begin{widetext}
\beq
    \Psi_n
    =
    \begin{cases}
        \Psi_L^+ u(k_+) e^{i k_+ (n-1)} + \Psi_L^- u(k_-) e^{i k_- (n-1)} & n < 1-\ell_\text{max}  \\
        \Psi_R^+ u(k_+) e^{i k_+ (n-N)} + \Psi_R^- u(k_-) e^{i k_- (n-N)} & n > N+\ell_\text{max}.\label{eq:scattering wavefunction general}
    \end{cases}
\eeq
\end{widetext}

The S matrix and scattering transfer matrix of the sample are $2\times 2$.  We would like to use the analytical result \eqref{eq:lloc to 2nd order}, but we face the difficulty that, due to the higher range of $U(t)$ in the disordered region, the scattering transfer matrix cannot be written in the factorized form \eqref{eq:scattering transfer matrix factorization} with i.i.d., $2\times 2$ scattering transfer matrices $\mathcal{T}_n$.  (Note that a factorized form with $2\times 2$ matrices may exist, but the disorder will generally not separate.  The dimer disorder model of Sec. \ref{sec:Phase disorder in dimers} illustrates this; though we worked in position space there, the scattering formulation would be similar.)

\begin{figure}[htb]
\subfloat[\label{subfig:scattering_setup}]{%
  \includegraphics[width=.6\columnwidth]{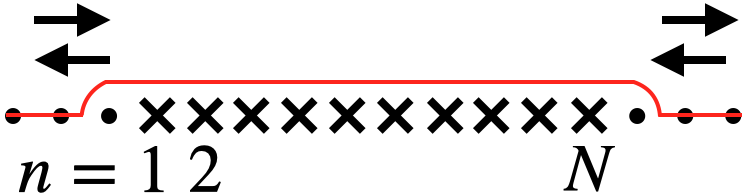}%
}\vfill
\subfloat[\label{subfig:spacer_setup}]{%
  \includegraphics[width=.6\columnwidth]{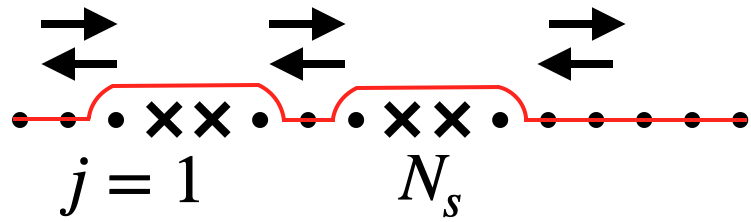}%
}
\caption{Schematic of our scattering setup.  Dots represent clean sites and crosses represent disordered sites.  The red line indicates the magnitude of the matrix elements of the disordered unitary $U(t)$ when at least one index is a disordered site.  (a) The original scattering problem, in which the disordered sites $n=1,\dots,N$ form a disordered sample within clean leads.   Here, $N=10$.  (b) A modified problem, in which we insert clean regions into the sample.  There are then $N_s$ disordered regions, each of which is a scatterer of the form of the original problem.  Here, $w_\text{block}=5$, $f=2/5$, and $N_s=2$.}
\end{figure}

We can overcome this difficulty in a modified family of scattering problems.  Before going into detail, let us state that the basic idea is to introduce clean regions, so the sample is replaced by an alternating sequence of disordered and clean regions (Fig. \ref{subfig:spacer_setup}).

Each disordered region is effectively a new sample, and in particular, has a $2\times 2$ scattering transfer matrix.  We thus obtain the factorized form \eqref{eq:scattering transfer matrix factorization}.  Our main task in the calculation below will be to show that we obtain the exponent $\nu=2$ no matter how small a fraction of the overall sample is made up of clean regions.  Thus, although we do not address the original problem, we do obtain the exponent in a family of problems that come arbitrarily close to the original problem.

We proceed to describe the setup in more detail.  Consider a sample of $N$ lattice sites divided into blocks of size $w_\text{block}$.  The total number of blocks, which we write as $N_s$, is then given by
\beq
    N_s= N / w_\text{block}.\label{eq:Ns}
\eeq
A parameter $f\in(0,1)$ denotes the fraction of disordered sites within each block: the first $f w_\text{block}$ sites of each block are disordered (we refer to this group of sites as a disordered region) and the remaining $(1-f)w_\text{block}$ sites are clean.  Here, a site $n$ is made clean by deleting the corresponding operator $\Theta_n(t)$ in Eq. \eqref{eq:Hdisorder}.  A unitary operator $U(f,t)$ is then defined by taking the time-ordered exponential of the Hamiltonian \eqref{eq:H_for_scattering} from $0$ to $t$.  Setting $f=1$ corresponds to a sample with no clean regions, which is the original scattering problem [$U(1,t)=U(t)$].

The idea is that the localization properties of $U(f,t)$ should be similar to those of $U(t)$, provided that $f\approx 1$.
Let us note that for $f \in (0,1)$, the modified unitary is generally \emph{not} a loop [$U(f,\td)\ne 1$]; however, our arguments below do not require the loop condition to hold for general $f$.

We proceed to define the localization length, given any fixed fraction $f\in(0,1)$ and any block size $w_\text{block}$.  At any $t\ne \td/2$, localization occurs and has the following effect: If we increase $N_s$ [with $N$ determined by Eq. \eqref{eq:Ns}], the typical transmission coefficient of the auxiliary problem decays exponentially in $N_s$.  We may write this decay as $T_\text{typ} \sim e^{-2 N_s w_\text{block}/\lloc(f,t)}$, where the factor of $w_\text{block}$ is included so $\lloc(f,t)$ is measured in units of the original lattice spacing.  (As usual, we are considering the localization length at some quasienergy.  The localization length also depends on $w_\text{block}$, but below we will take $w_\text{block}$ to be determined by $f$.)

We now show that at any fixed fraction $f\in(0,1)$, the exponent $\nu=2$ is obtained for $\lloc(f,t)$.  Given $f$, we first fix a value for $w_\text{block}$ that is large enough to prevent any neighboring disordered regions from overlapping.  In particular, we require
\beq
    (1-f)w_\text{block} > 2\ell_\text{max}.\label{eq:wblock requirement}
\eeq

We can then write the scattering wavefunction as a linear combination of Bloch waves in all of the clean regions.  We follow the same phase convention as in Eq. \eqref{eq:scattering wavefunction general}.  To do this, it is convenient to write the leftmost and rightmost site in the $j$th disordered region as
\bseq
\begin{align}
    n_{L,j}&= w_\text{block}(j-1)+1,\\
    n_{R,j}&= n_{L,j} + f w_\text{block} -1.
\end{align}
\eseq
Then we may write, for any $j=1,\dots,N_s$,
\begin{widetext}
\beq
    \Psi_n =
    \begin{cases}
        \Psi_{L,j}^+ u(k_+)e^{i k_+(n-n_{L,j})}+ \Psi_{L,j}^- u(k_-)e^{i k_-(n-n_{L,j})} & n_{R,j-1} +\ell_\text{max} < n < n_{L,j} - \ell_\text{max}\\
        \Psi_{R,j}^+ u(k_+)e^{i k_+(n-n_{R,j})}+ \Psi_{R,j}^- u(k_-)e^{i k_-(n-n_{R,j})} & n_{R,j} +\ell_\text{max} < n < n_{L,j+1} - \ell_\text{max},
    \end{cases}\label{eq:scattering wavefunction in spaces}
\eeq
where $n_{R,0}\equiv -\infty$ and $n_{L,N_s+1}\equiv \infty$.  Comparing to Eq. \eqref{eq:scattering wavefunction general}, we get $\Psi_L^\pm = \Psi_{L,1}^\pm$ and $\Psi_R^\pm = \Psi_{R,N_s}^\pm$.  Also, comparing $j$ to $j+1$ in Eq. \eqref{eq:scattering wavefunction in spaces}, we get
\beq
    \bpmat
        \Psi_{L,j+1}^+ \\
        \Psi_{L,j+1}^-
    \epmat
    =
    \bpmat
        e^{ik_+[(1-f)w_\text{block}+1]} & 0\\
        0 & e^{ik_-[(1-f)w_\text{block}+1]} 
    \epmat
    \bpmat
        \Psi_{R,j}^+ \\
        \Psi_{R,j}^-
    \epmat,
\eeq
\end{widetext}
where we have noted that $n_{L,j+1}- n_{R,j}= (1-f)w_\text{block}+1$.

The scattering transfer matrix that relates $\Psi_{L,j}^\pm$ to $\Psi_{R,j}^\pm$ is just the scattering transfer matrix of the original problem with a different number of sites ($f w_\text{block}$ sites instead of $N$).  Let us write this matrix as $\hat{\mathcal{T}}_j$, bearing in mind that the subscript $j$ stands for dependence on all of the disorder variables that appear in the $j$th disordered region.  We then have
\beq
    \bpmat
        \Psi_{R,j}^+\\
        \Psi_{R,j}^-
    \epmat
    = \hat{\mathcal{T}}_j 
    \bpmat
        \Psi_{L,j}^+\\
        \Psi_{L,j}^-
    \epmat.
\eeq

Collecting the last several equations, we obtain the desired factorization \eqref{eq:scattering transfer matrix factorization} with
\beq
    \mathcal{T}_j = \hat{\mathcal{T}}_j \bpmat
        e^{ik_+[(1-f)w_\text{block}+1]} & 0\\
        0 & e^{ik_-[(1-f)w_\text{block}+1]}\label{eq:scattering transfer matrix for a block} 
    \epmat,
\eeq
in which the matrix multiplying on the right only introduces unimportant phase factors.  By construction, the disorder dependence in the matrices $\mathcal{T}_1,\dots,\mathcal{T}_{N_s}$ is i.i.d. because the disordered regions have no overlap.  Thus, Eq. \eqref{eq:lloc to 2nd order} applies.  In particular, in units of the original lattice spacing, we have
\begin{widetext}
\beq
    2 /\lloc(f,t) = \frac{1}{w_\text{block}} \left( \langle R_j\rangle_j - 2 \text{Re}\left[ \frac{\langle r_j \rangle_j \langle r_j'\rangle_j}{1+ \langle r_j r_j'/R_j\rangle_j }\right] \right)+O(|r_j|^3).
\eeq
\end{widetext}

To complete the calculation, we next argue that the reflection amplitude of an individual block satisfies $|r_j| \sim \dt$.  The essential point is to show $|\hat{r}_j|\sim \dt$, where $\hat{r}_j$ is the reflection amplitude associated with $\hat{\mathcal{T}}_j$; once this is shown, we can easily account for the time dependence introduced by the phase factors that appear in Eq. \eqref{eq:scattering transfer matrix for a block}.  Indeed, from Eq. \eqref{eq:scattering transfer matrix for a block} and the general parametrization \eqref{eq:scattering transfer matrix general parametrization}, we get
\bseq
\begin{align}
    r_j &= \hat{r}_j e^{i(k_+ - k_-)[(1-f)w_\text{block} + 1 ]},\label{eq:r in terms of rhat}\\
    r_j' &= \hat{r}_j'.
\end{align}
\eseq
From \eqref{eq:wblock requirement}, we see that it suffices to take $w_\text{block}$ to be, e.g., $3\ell_\text{max}/(1-f)$; in particular, $(1-f)w_\text{block}$ is then independent of $t$.  The momenta $k_\pm$ depend on $t$ in a regular way, i.e., $k_\pm = k_\pm^{(0)} + O(\dt)$.  Thus, we conclude that $|r_j| \sim \dt$ will follow from $|\hat{r}_j|\sim \dt$.  

To argue that $|\hat{r}_j|\sim\dt$, we begin by emphasizing the difference between two limits of the original scattering problem ($f=1$): (1) If we fix $\dt>0$ and send the sample size $N$ to infinity, then the magnitude of the sample reflection amplitude goes to unity due to localization ($|r_{1\dots N}|\to 1$).  This is the regime we are interested in.  However, for the purpose of treating each disordered region as a version of the original problem, it is more useful to consider the following limit.  (2) If we instead fix the system size $N$, then we have some expansion of $r_{1\dots N}$ in powers of $\dt$.  We claim that this expansion generically starts at first order, i.e., $r_{1\dots N} \sim \dt$.

The key point in our argument for (2) is our statement in Sec. \ref{sec:Universal localization-delocalization transition}, namely, that there are delocalized states throughout the quasienergy spectrum at $\dt=0$.  In the scattering framework, a state is delocalized if and only if its reflection amplitude vanishes; that is, the disorder can only appear in the transmission phases.  Thus, at each quasienergy, there must be a state with $r_{1\dots N}\rvert_{\dt = 0}=0$.
This shows that there is no zeroth order term in the expansion of $r_{1\dots N}$ in $\dt$.

Generically, the first non-vanishing order should be linear ($\dt$), essentially because $r_{1\dots N}$ is an amplitude.  Indeed, recall that in static scattering problems, the reflection amplitudes generically start at linear order in the scattering potential, in the Born approximation.  In Appendix \ref{sec:Reflection amplitude starts at linear order}, we use the discrete-time scattering formalism, as developed in Ref. \cite{BisioScattering2021}, to verify that $r_{1\dots N}$ indeed generically starts at linear order in $\dt$.

To obtain $|\hat{r}_n|\sim \dt$, we apply (2) to each disordered region.  The size of each disordered region ($f w_\text{block}$) is determined entirely by $f$ and $\ell_\text{max}$; in particular, it is independent of $t$.  Thus, we generically have $|\hat{r}_n|\sim \dt$.  From Eqs . \eqref{eq:lloc to 2nd order}, we thus obtain $1/\lloc(f,t)\sim (\dt)^2$, i.e., $\nu=2$ holds for any $f\in(0,1)$.

Throughout the above argument, we have assumed $U(t)$ to be strictly local [Eq. \eqref{eq:U(t) strictly local}].  However, a generic drive is only exponentially local, i.e., for sufficiently large $|n-n'|$ we have
\beq
    |U_{nn'}(t)| \le C(t) e^{- |n-n'|/ \xi(t)}\label{eq:U(t) exponentially local}
\eeq
for some constants $C(t)$ and $\xi(t)$ that are independent of $n,n'$.  We assume that we can replace $C(t)\to C$ and $\xi(t)\to\xi$ for some $t$-independent constants $C$ and $\xi$.  Then, we repeat the above arguments with $\ell_\text{max}$ fixed to any particular value satisfying $\ell_\text{max}\gg \xi$.  Although the disordered unitary then differs slightly from the clean unitary even in the clean regions, the difference can be made arbitrarily small.

Finally, let us state what would need to be shown to obtain the exponent $\nu=2$ in the original scattering problem.  Our argument above yields the small $\dt$ expansion $1/\lloc(f,t) = \frac{A(f)}{w_\text{block}}(\dt)^2$ for some unknown function $A(f)$.  Noting from \eqref{eq:wblock requirement} that $w_\text{block}\to\infty$ as $f\to1$, we see that $\nu=2$ can be obtained in the limit $f\to1$ given the assumption that $A(f)\sim w_\text{block}$.

\paragraph{Generalization of the clean drive.}

We next generalize $U_\text{clean}(t)$ beyond the particular case of the SSH-type drive.  Here we note that the precise forms of the Bloch function $u(k)$ and of the two momenta $k_{\pm}$ were not important in our calculation above; the essential point was that the S matrix for the sample was $2\times 2$.  We can therefore generalize our calculation to include any quasienergy at which $U_\text{clean}(t)$ has two-fold degeneracy for all $t$ in some neighborhood of $\td/2$.

\begin{figure}[htb]
\centering
    \includegraphics[width=\columnwidth]{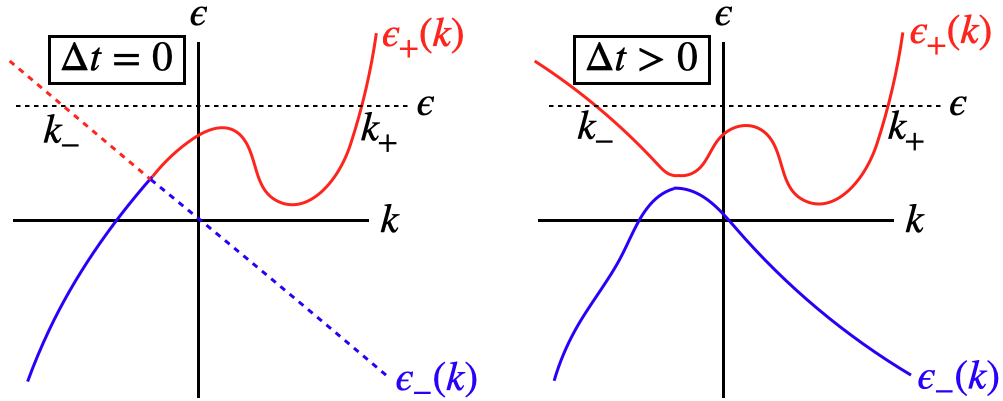}
    \caption{Illustration of the band structure of a more general two-band drive $U_\text{clean}(t)$.  A particular quasienergy $\epsilon$ is fixed, while $\dt$ is varied.  Left: The midpoint of the drive ($\dt=0$).  The solid line is $\epsilon_A(k)$ and the dashed line is $\epsilon_B(k)$; they reflect the winding numbers ($1$ and $-1$) of $U_{AA}(k)$ and $U_{BB}(k)$.  Though some quasienergies have higher degeneracy, there is a range of quasienergies that have two-fold degeneracy, with one right mover ($k_+$) and one left mover ($k_-$).  Right: Small, non-zero $\dt$.  The values of $k_+$ and $k_-$ depend on $\dt$, but there remains a similar range of quasienergies with twofold degeneracy.}\label{fig:band_structure}
\end{figure}

Let us consider the case that $U_\text{clean}(t)$ has two bands, which we write as $\epsilon_{\pm}(t,k)$.  In momentum space, the block-diagonal form \eqref{eq:Umidpoint as UAA+UBB} becomes
\beq
    U_\text{clean}(\td/2,k) = 
    \bpmat
    U_{AA}(k) & 0\\
    0 & U_{BB}(k)
    \epmat,
\eeq
where, by assumption, $U_{AA}(k)$ and $U_{BB}(k)$ have winding numbers $1$ and $-1$, respectively.  The corresponding quasienergy spectra $\epsilon_{A}(k)$ and $\epsilon_{B}(k)$ must go from $\epsilon= - 2\pi/\td$ to $2\pi/\td$, but need not do so monotonically.   Example spectra are sketched in Fig. \ref{fig:band_structure}.

For $t$ close to but not equal to $\td/2$, gaps open at all band crossings.  We define $\epsilon_+(k)$ and $\epsilon_-(k)$ to be the upper and lower bands, respectively, and we focus for now on the midpoint of the drive, at which the bands $\epsilon_\pm(k)$ are defined by taking the limit as $t\to\td/2$ .  [Note that the bands $\epsilon_{\pm}(k)$ at the midpoint are not the same as $\epsilon_A(k)$ and $\epsilon_B(k)$.]  As is clear from Fig. \ref{fig:band_structure}, there is generically some \emph{range} of quasienergies $\epsilon$ that have a two-fold degeneracy with the two associated momenta moving in opposite directions.  It is for $\epsilon$ in this range that our argument straightforwardly applies.

Let us explain this in more detail.  For $\epsilon$ in some range, we have $\epsilon= \epsilon_+(k_-)=\epsilon_+(k_+)$, $v_g(k_+)>0$, and $v_g(k_-)<0$, where $v_g(k)=\pd\epsilon_+(k)/\pd k$ is the group velocity.  (Here we assume for definiteness that $\epsilon$ lies in the upper band.)  Although we have considered $\epsilon_\pm(k)$ at the midpoint ($t= \td/2$), allowing $t$ to vary slightly from the midpoint does not change the qualitative features; there remains a similar range of quasienergies with the desired two-fold degeneracy.  For quasienegies in this range, and for $t$ sufficiently close to the midpoint, the eigenstates of $U_\text{clean}(t)$ are linear combinations of two (time-dependent) Bloch waves.  We have thus brought our localization problem, for a more general class of topologically nontrivial drives, into the framework of scattering with $2\times 2$ matrices, at least for some range of quasienergies.

We now present the scattering argument for the exponent $\nu=2$ in this more general setting.  A complication arises, both in the basic setup of the scattering problem and in our argument about inserting clean regions: although by assumption $U_\text{clean}(t)$ has two bands, it may have eigenstates with complex momenta when confined either to a half-line or to a finite region.  These eigenstates grow or decay exponentially in position space; schematically, $\bra{n} \ket{\Psi_\epsilon(t)} \sim e^{\pm n/ r_\epsilon(t)}$ for some decay length $r_\epsilon(t)$.  [We emphasize that this length scale is distinct from the operator decay length $\xi(t)$ that appears in Eq. \eqref{eq:U(t) exponentially local}.  Even a strictly local $U_\text{clean}(t)$ can have exponentially decaying eigenstates, although the SSH-type drive of Sec. \ref{sec:Example drives with the universal exponent} does not.]  When we introduce disorder into sites $n=1,\dots, N$, the edge sites ($n=1$ and $N$) can then be regarded as endpoints of half-lines that extend into the leads.  This implies that the eigenstate wavefunction in the leads is generally not a linear combination solely of Bloch waves [even at distances greater than $\xi(t)$ away from the sample]; there can also be contributions from modes that decay as $n\to\pm\infty$.  The solution is simple: as usual in scattering theory, the expansion in terms of Bloch waves [Eq. \eqref{eq:scattering wavefunction general}] holds for $n\to\pm\infty$.

Similarly, we must note that the exponentially growing and decaying eigenstates of $U_\text{clean}(t)$ can appear in each inserted clean region [even at distances greater than $\xi(t)$ from the nearest disordered region].  To deal with this, we assume that the decay lengths of the eigenstates of $U_\text{clean}(t)$ can be uniformly bounded, in the sense that there is a fixed distance $r_\epsilon$, independent of $t$, for which $r_\epsilon(t)< r_\epsilon$ for all $t$ (in some neighborhood of $\td/2$).  We choose the constant $\ell_\text{max}$ to be much greater than $r_\epsilon$, i.e., $\ell_\text{max}$ satisfies
\beq
    \ell_\text{max} \gg \text{max}\{ \xi,r_\epsilon\}.
\eeq
Then, the expansions in terms of Bloch waves---both the basic setup [Eq. \eqref{eq:scattering wavefunction general}] and the more general expansions within each clean region [Eq. \eqref{eq:scattering wavefunction in spaces}]---hold as written (up to error that can be decreased arbitrarily by increasing $\ell_\text{max}$).  From this point onward, our argument for the exponent $\nu=2$ proceeds exactly as in the case that $U_\text{clean}(t)$ was taken to be the SSH-type drive.

Thus, we have obtained the exponent $\nu=2$ provided that we consider a quasienergy $\epsilon$ at which the clean drive (evaluated at the midpoint) has twofold degeneracy.  In the generic two-band case, there is a range of $\epsilon$ for which this condition holds.

\section{Conclusion}\label{sec:Conclusion}
We considered the class AIII of Floquet drives in one spatial dimension, focusing on drives with topological index equal to one.  We argued that these drives exhibit a localization-delocalization transition, with the localization length diverging throughout the quasienergy spectrum at a particular time (the midpoint of the loop part of the drive).  We obtained the delocalization exponent $\nu=2$ explicitly in a simple model, checked it numerically in more complicated models, and provided an analytical argument for a class of models (based on some plausible assumptions).  We expect that some assumptions of the analytical argument are not essential; for instance, the restriction to twofold degeneracy in the clean drive.  Also, we expect that the same exponent holds for any nonzero value of the topological index.

A basic assumption throughout this work is the chiral symmetry, which relates the Hamiltonian to itself at different times.  Since we focus on disordered systems, this symmetry property could be difficult to achieve in a real system.  To address this, it would be interesting to see if the exponent persists even if the symmetry holds only in a disorder-averaged sense rather than holding for every disorder realization \cite{FulgaStatistical2014}.  Alternatively, the setup we discuss at the end of Sec. \ref{sec:Exceptions to the universal exponent, and further generality}, which yields the same exponent without appeal to a time-dependent symmetry constraint, could be explored further.

Another natural follow-up to this work would be to consider the effect of interactions.  The SSH-type drive we have considered readily extends to an interacting spin model; note also that the interacting generalization of the flow index is known from Ref. \cite{GrossIndex2012}.  The interacting generalization of our work may provide an interesting example of a localization-delocalization transition in a Floquet-many-body-localized setting.

\acknowledgments{We thank Victor Gurarie, Curt von Keyserlingk, and Shivaji Sondhi for useful discussions on related topics.  This work used computational and storage services associated with the Hoffman2 Shared Cluster provided by UCLA Institute for Digital Research and Education’s Research Technology Group.  A.B.C., P.S., and R.R. acknowledge financial support from the University of California Laboratory Fees Research Program funded by the UC Office of the President (UCOP), grant number LFR-20-653926.  A.B.C acknowledges financial support from the Joseph P. Rudnick Prize Postdoctoral Fellowship (UCLA).  P.S. acknowledges financial support from the Center for Quantum Science and Engineering Fellowship (UCLA) and the Bhaumik Graduate Fellowship (UCLA). P.S. acknowledges the support of NNSA for the U.S. DoE at LANL under Contract No. DE-AC52-06NA25396, and Laboratory Directed Research and Development (LDRD) for support through 20240032DR. LANL is managed by Triad National Security, LLC, for the National Nuclear Security Administration of the U.S. DOE under contract 89233218CNA000001. Research presented in this article was also supported by the National Security Education Center (NSEC) Informational Science and Technology Institute (ISTI) using the Laboratory Directed Research and Development program of Los Alamos National Laboratory project number 20240479CR-IST.}

\appendix
\begin{widetext}
\section{Further details for the SSH-type drive with on-site disorder}\label{sec:Further details for the SSH-type drive with on-site disorder}

\subsection{Mapping to discrete-time quantum walk}\label{sec:Mapping to discrete-time quantum walk}

In the main text, we showed that the eigenstate problem [Eq. \eqref{eq:eigenstate eq}] becomes Eqs. \eqref{eq:SchrodA extended} and \eqref{eq:SchrodB extended} in position space.  Here, we show that these equations are equivalent to a special case of the generalized discrete-time quantum walk studied in Ref. \cite{VakulchykAnderson2017}.  We start by reviewing the setup for the discrete-time quantum walk \cite{VakulchykAnderson2017}.

The discrete-time quantum walk occurs on an infinite chain with a two-component  ``spin'' degree of freedom ($\uparrow$, $\downarrow$) at each site; a general state is written as $\ket{\Psi}=\sum_n (\Psi_{n,\uparrow}\ket{n,\uparrow} + \Psi_{n,\downarrow}\ket{n,\downarrow}$.  A single time step is implemented by a unitary operator $\hat{U}= \hat{S}\hat{U}_\text{coin}$, where the shift $\hat{S}$ moves up (down) spins one unit to the right (left) and where $\hat{U}_\text{coin}$ acts on the spin degree of freedom at each site $n$ as a unitary $2\times 2$ matrix $U_{\text{coin},n}$.  The matrix $U_{\text{coin},n}$ is parameterized by $\varphi_n, \varphi_{1,n},\varphi_{2,n}$, and $\theta_n$ (in the same way as in Eq. (1) of Ref. \cite{VakulchykAnderson2017}).

The eigenstate equation for the discrete-time quantum walk problem is $\hat{U}\ket{\Psi} = e^{-i\omega}\ket{\Psi}$.  In position space, the eigenstate equations become Eqs. (H1a) and (H2b) of Ref. \cite{CulverScattering2024}.  Our Eqs. \eqref{eq:SchrodA extended} and \eqref{eq:SchrodB extended} are in fact special cases of (H1a) and (H2b) [with $n$ relabelled as $n-1$ in (H1a) and as $n+1$ in (H2b)] with the following choice of parameters:
\bseq
\begin{align}
    \varphi_n &= \frac{1}{2}(\phi_{n,A}+\phi_{n+1,B}),\label{eq:varphin}\\
    \varphi_{1,n}&=\frac{1}{2}(\phi_{n,A}-\phi_{n+1,B}),\\
    \varphi_{2,n}&= -\frac{1}{2}(\phi_{n,A} - \phi_{n+1,B}) + \pi/2,\\
    \theta_n&= v_n \dt\label{eq:thetan},
\end{align}
\eseq
and with the following correspondence of the basic variables:
\beq
    \Psi_{n,\uparrow} \leftrightarrow \Psi_{n,A},\ \Psi_{n-1,\downarrow} \leftrightarrow \Psi_{n,B},\ \omega\leftrightarrow \epsilon t.\label{eq:correspondence to DTQW}
\eeq

Using this mapping, we can reproduce some of the results below by translating results from the discrete-time quantum walk case.  However, for completeness we present the calculations directly in the case of the SSH-type drive.

\subsection{Solution of the clean model}\label{sec:Solution of the clean model}

By Bloch's theorem, a wavefunction that solves the eigenstate equation \eqref{eq:eigenstate eq} with some quasienergy $\epsilon(k)$ can be written as
\beq
    \Psi_n
    = 
    u(k)
    e^{i k n},
\eeq
where the Bloch function $u(k)$ is a two-component spinor in the sublattice index:
\beq
    u(k) \equiv 
    \bpmat
        u_{A}(k) \\
        u_{B}(k)
    \epmat. \label{eq:Bloch wavefn}
\eeq

We proceed to solve for the quasienergies and Bloch functions.  From now on, we assume that are in the middle section of the drive ($\td/ < t < 3\pi/2$).  We remove the disorder from Eqs. \eqref{eq:SchrodA phase disorder} and \eqref{eq:SchrodB phase disorder} (by setting all $\phi_{n,A}=\phi_{n,B}=0$) to obtain
\bseq
\begin{align}
        \cos\left(\dt \right)\ \Psi_{n-1,A} + i \sin( \dt)\  \Psi_{n,B}  &= e^{-i \epsilon t} \Psi_{n,A},\\
        i \sin (\dt)\ \Psi_{n,A} + \cos(\dt)\  \Psi_{n+1,B} &= e^{-i \epsilon t} \Psi_{n,B}.
\end{align}
\eseq
Substituting in the plane wave form given above, we obtain
\beq
    \bpmat
        e^{-i k } \cos(\dt) & i \sin (\dt)\\
        i \sin (\dt) & e^{i k }\cos (\dt)
    \epmat
    u(k) = e^{-i \epsilon(k) t} u(k),
\eeq
which we then solve to find the two bands of the model (denoted $+$ and $-$).  The quasienergies of the two bands are defined (modulo $2\pi/t$) by
\beq
    e^{-i \epsilon_{\pm}(k) t}  = \cos(\dt) \cos k \mp i \sqrt{1- \cos^2 \dt \cos^2 k },
\eeq
and the unit-normalized Bloch spinors are
\beq
    u_{\pm}(k) = \frac{\sgn k}{\mathcal{N}_{\pm}(k) }
    \bpmat
        \sin(\dt) \\
        \cos(\dt) \sin k \mp \sqrt{1- \cos^2 \dt \cos^2 k}
    \epmat, \label{eq:u_defn}
\eeq
where the prefactor of $\sgn k$ is chosen for convenience and where $\mathcal{N}_\pm(k)$ is a normalization factor:
\beq
    \mathcal{N}_\pm(k) = \sqrt{2 \left( 1- \cos^2 \dt\cos^2 k \mp \cos(\dt) \sin k \sqrt{1- \cos^2 \dt \cos^2 k} \right)} .
\eeq
The first Brilluoin zone is $|k|<\pi$ and $\epsilon_\pm(k)$ is fixed by requiring $|\epsilon_\pm(k) t|<\pi $.  The resulting two bands are shown in Fig \ref{fig:spectrum}.

\subsection{Calculation of the localization length}\label{sec:Calculation of the localization length}

\subsubsection{Position space calculation}\label{sec:Position space calculation}
Here we obtain Eq. \eqref{eq:lloc model drive phase and bond disorder} using a result from Schrader \emph{et al}. \cite{SchraderPerturbative2004}.  For a summary of their setup in a notation similar to that used in this paper, and also for a calculation that includes that below as a special case (using the mapping from Appendix \ref{sec:Mapping to discrete-time quantum walk}), see Appendix B of Ref. \cite{CulverScattering2024}.

As we point out in the main text, the (position-space) transfer matrix $\mathcal{M}_n$ satisfies $\mathcal{M}_n^\dagger \sigma^z \mathcal{M}_n=\sigma^z$.  Reference  \cite{SchraderPerturbative2004} instead considers matrices satisfying the same condition with $\sigma^z$ replaced by $\sigma^y$; however, as they point out, the two types of matrix are isomorphic.  We therefore start by defining a matrix $K_n$ that is isomorphic to $\mathcal{M}_n$: in particular, $K_n = C \mathcal{M}_n C^{-1}$, where the unitary matrix $C$ is given by \cite{SchraderPerturbative2004}
\beq 
    C \equiv \frac{1}{\sqrt{2}}
    \bpmat
    i & i \\
    1 & -1
    \epmat.
\eeq
Then $K_n$ satisfies the $\sigma^y$ condition.

Equation~\eqref{eq:Lyapunov exponent definition} can be written in terms of the $K_n$ matrices as
\beq
    \gamma(t) = \lim_{N\to\infty} \frac{1}{N}\langle \ln || K_N\dots K_1 || \rangle_{1\dots N},\label{eq:Lyapunov exponent Schrader form}
\eeq
Note that in passing from Eq. \eqref{eq:Lyapunov exponent definition} to Eq. \eqref{eq:Lyapunov exponent Schrader form}, we dropped boundary terms that make no contribution in the limit $N\to\infty$.

Reference \cite{SchraderPerturbative2004} calculates \eqref{eq:Lyapunov exponent Schrader form} to leading order in a small parameter $\lambda$ which is given in our case by $\lambda=\dt$.  
To use the result from Ref. \cite{SchraderPerturbative2004}, we must verify that $K_n$ (there denoted $T_{\lambda,\sigma}$) satisfies the necessary properties.  First, we check that the matrices $K_n$ commute with each other at $\dt=0$.  At the point $\dt=0$, we have
\beq
    \mathcal{M}_n =
    \bpmat
        e^{i (\pi \epsilon + \phi_{n,A})} & 0\\
         0 & e^{-i (\pi \epsilon + \phi_{n+1,B})}
    \epmat,
\eeq
which implies in particular that $[\mathcal{M}_n, \mathcal{M}_{n'}]\rvert_{\dt=0} =0$; then the same is true for the $K_n$ matrices, by linearity.  Second, we must check the trace condition $|\text{Tr} K_n\rvert_{\dt=0}|<2$, which in our case reduces to
\beq
    |e^{i(\pi\epsilon + \phi_{n,A})} + e^{-i(\pi\epsilon + \phi_{n+1,B}}|< 2.
\eeq
At worst, the left-hand side can equal $2$, but only in a fine-tuned case; generically, the inequality is satisfied.

We then define
\bseq
\begin{align}
    \eta_n &= \pi\epsilon+\frac{1}{2}(\phi_{n,A} + \phi_{n+1,B}),\label{eq:eta}\\
    \xi_n &= \frac{1}{2}(\phi_{n,A} + \phi_{n+1,B}),\\
    P_n &= 
    \bpmat
        v_n \sin(\phi_{n,A}+\pi\epsilon) & v_n \cos(\phi_{n,A}+\pi\epsilon)-\epsilon \\ v_n \cos(\phi_{n,A}+\pi\epsilon)+\epsilon & -v_n \sin(\phi_{n,A}+\pi\epsilon)
    \epmat.
\end{align}
\eseq
Note that $P_n$ is real and traceless.  We then have, by a straightforward calculation, a parametrization of the form of Eq. (8) from Ref. \cite{SchraderPerturbative2004}:
\beq
    K_n = e^{i\xi_n} R_{\eta_n} \left( \mathbb{I} + \dt P_n \right) +  O((\dt)^2),\label{eq:decomposition}
\eeq
where $R_{\eta_n}$ is the $2\times 2$ matrix that rotates by the angle $\eta_n$.  [Eq. (8) in Ref. \cite{SchraderPerturbative2004} also includes a matrix $M \in \text{SL}(2,\mathbb{R})$, which is in our case the identity matrix, and another matrix $Q_n$ that is not needed for our purposes.]  The constant $\beta_n$ defined (in a slightly different notation) by Eq. (9) from Ref. \cite{SchraderPerturbative2004} is then found to be
\beq
    \beta_n = -i v_n e^{i(\pi \epsilon + \phi_{n,A})}.\label{eq:beta}
\eeq
Finally, substituting Eqs. \eqref{eq:eta} and \eqref{eq:beta} into Eq. (22) of Ref. \cite{SchraderPerturbative2004} yields Eq. \eqref{eq:lloc model drive phase and bond disorder} from the main text.

\subsubsection{Scattering calculation}\label{sec:Scattering calculation}
In Sec. \ref{sec:Scattering calculation for the SSH-type drive with on-site disorder}, we obtained the inverse localization length at leading order in $\dt$ [Eq. \eqref{eq:lloc model drive phase and bond disorder}] by applying the scattering formula \eqref{eq:lloc to 2nd order} to the auxiliary problem defined by Eq. \eqref{eq:auxiliary problem SSH-type drive}.  This required us to show that the auxiliary problem has the same localization length as the original problem.  In this section, we apply Eq. \eqref{eq:lloc to 2nd order} directly to the original problem, yielding the same result \eqref{eq:lloc model drive phase and bond disorder}.

Without loss of generality, we fix $\epsilon>0$ and consider $\dt>0$.  The scattering momentum $k>0$ and the Bloch function $u(k)$ vary with $t$.  Expanding in $\dt$, we readily obtain
\bseq
\begin{align}
    k&= \pi\epsilon + O((\dt)^2),\\
    u(k) &= 
    \bpmat
        1\\
        -\frac{1}{2}(\csc k) \dt
    \epmat
    + O((\dt)^2)\\
    u(-k) &= 
    \bpmat
        -\frac{1}{2}(\csc k) \dt\\
        1
    \epmat
    + O((\dt)^2).
\end{align}
\eseq
Then, from Eqs. \eqref{eq:transfer matrix phase and bond disorder}, \eqref{eq:matLambda}, and \eqref{eq:matTn}, we calculate the scattering transfer matrix $\mathcal{T}_n$ as an expansion in $\dt$.  We only need the reflection amplitudes at linear order in $\dt$, and they can be obtained straightforwardly using the general parametrization \eqref{eq:scattering transfer matrix general parametrization}:
\bseq
\begin{align}
    r_n &= \frac{1}{2} \left(\csc k - e^{i(2k + \phi_{n,A} + \phi_{n+1,B})}\csc k + 2i v_n e^{i (k+\phi_{n,A})} \right) \dt + O((\dt))^2,\\
    r_n' &= \frac{1}{2} \left(\csc k - e^{i(2k + \phi_{n,A} + \phi_{n+1,B})}\csc k + 2i v_n e^{i (k+\phi_{n+1,B})} \right) \dt + O((\dt))^2.
\end{align}
\eseq
Substitution into Eq. \eqref{eq:lloc to 2nd order} yields Eq. \eqref{eq:lloc model drive phase and bond disorder} from the main text once again.

\subsubsection{Full phase disorder}\label{sec:Full phase disorder}
To obtain Eq. \eqref{eq:Lloc any dt model drive}, we start by recalling a result from Ref. \cite{AndersonNew1980}.  Consider a general scattering problem in which we have the factorization \eqref{eq:scattering transfer matrix factorization} with $N_s=N$ (hence we will write $\mathcal{T}_j\equiv \mathcal{T}_n$).  Let $\mathcal{T}_n$ be parametrized by $r_n$, $r_n'$, $t_n$, and $t_n'$, as in Eq. \eqref{eq:scattering transfer matrix general parametrization}.  If, at sufficiently large system size $N$, we have the following phase uniformity condition:
\beq
    \text{Arg}[ r_{1\dots N}'r_{N+1}] \text{ is distributed uniformly in }[-\pi,\pi], \text{independently of }|r_{N+1}|,\label{eq:uniform phase condition}
\eeq
then Ref. \cite{AndersonNew1980} shows
\beq
    \frac{2}{\lloc} = \langle -\ln T_n\rangle_n,\label{eq:lloc AndersonNew1980}
\eeq
where $T_n = |t_n|^2 =|t_n'|^2$ is the transmission coefficient and where the disorder average is taken over any site $n=1,\dots,N$.  Since $\text{Arg}[ r_{1\dots N}'r_{N+1}]= \text{Arg}[ r_{1\dots N}'] + \text{Arg}[r_{N+1}]$, one simple case in which the condition \eqref{eq:uniform phase condition} holds is the following: $\text{Arg}[r_n]$ is distributed uniformly in $[-\pi,\pi]$, independently of $|r_n|$. 

We now apply this result to the SSH-type drive with on-site phase disorder.  (We do not consider bond disorder because this was only defined for times $t$ near the midpoint, whereas here our concern is to find an answer for all $t$.)  It is simplest to work with the auxiliary problem defined by Eq. \eqref{eq:auxiliary problem SSH-type drive}.  Setting $v_n=1$ in Eq. \eqref{eq:transfer matrix phase and bond disorder}, we get $\text{Arg}[\tilde{r}_n] = \epsilon t + \phi_{n,A} + \frac{\pi}{2}\text{sgn} \dt$ and $|\tilde{r}_n|= |\sin \dt|$.  Thus, if $\phi_{n,A}$ is uniformly distributed in $[-\pi,\pi]$, then the condition \eqref{eq:uniform phase condition} holds, and \eqref{eq:lloc AndersonNew1980} yields Eq. \eqref{eq:Lloc any dt model drive} from the main text.

\section{Transfer matrix for the SSH-type drive with dimer disorder}\label{sec:Transfer matrix for the SSH-type drive with dimer disorder}
Here we derive the transfer matrix relation \eqref{eq:transfer matrix relation dimer disorder} and the explicit expression for the transfer matrix $\mathcal{M}_{n,A}$ for the $A$ sites.  To simplify the notation, we relabel the four sites $\ket{2n-1,A}$, $\ket{2n-1,B}$, $\ket{2n,A}$ and $\ket{2n,B}$ as $\ket{n,a}$,$\ket{n,b}$, $\ket{n,c}$ and $\ket{n,d}$, respectively.  We write the disordered unitary matrices $U_{n,\alpha}$ ($\alpha=A$ or $B$) as
\beq
    U_{n,A} =
    \begin{pmatrix}
        U_n^{aa} & U_n^{ac} \\
        U_n^{ca} & U_n^{cc}
    \end{pmatrix}
    ,\ 
    U_{n,B} =
    \begin{pmatrix}
        U_n^{bb} & U_n^{bd} \\
        U_n^{db} & U_n^{dd}
    \end{pmatrix}
    .
\eeq

It is straightforward to show that the eigenstate equation \eqref{eq:eigenstate eq} becomes the following four equations in position space:

\bseq
\begin{align}
     e^{- i  \epsilon  t}\Psi_{n,a} &= \cos (\dt) (U_{n-1}^{ca}  \Psi_{n-1,a} + U_{n-1}^{cc} \Psi_{n-1,c})   + i \sin (\dt)  (U_n^{bb} \Psi_{n,b}+ U_n^{bd} \Psi_{n,d} )   ,\\
    e^{- i  \epsilon  t}\Psi_{n,b} &= i \sin (\dt)(U_n^{aa} \Psi_{n,a}  + U_n^{ac} \Psi_{n,c})   + \cos (\dt) ( U_n^{db} \Psi_{n,b}+ U_n^{dd} \Psi_{n,d}),\\
    e^{- i  \epsilon  t}\Psi_{n,c}&= \cos (\dt) (U_n^{aa} \Psi_{n,a}  + U_n^{ac} \Psi_{n,c}) + i \sin (\dt) (U_n^{db} \Psi_{n,b} + U_n^{dd} \Psi_{n,d}), \\
    e^{- i  \epsilon  t} \Psi_{n,d}&= i \sin (\dt)(U_{n}^{ca}  \Psi_{n,a} + U_{n}^{cc} \Psi_{n,c})  + \cos (\dt) (U_{n+1}^{bb} \Psi_{n+1,b}+ U_{n+1}^{bd} \Psi_{n+1,d} ). 
\end{align}
\eseq
We use the second and third equations to eliminate $\Psi_{n,b}$ and $\Psi_{n,c}$ [this is expressed by Eq. \eqref{eq:Mprime}].  From the first and fourth equations, we then obtain
\beq
    \bpmat
        \Psi_{n+1,a}\\
        \Psi_{n+1,d}
    \epmat
    = \mathcal{M}_{n,\text{dimer}}
    \bpmat
        \Psi_{n,a}\\
        \Psi_{n,d}
    \epmat,
\eeq
where the transfer matrix $\mathcal{M}_{n,\text{dimer}}$ is given below.  Note that this is the same as Eq. \eqref{eq:transfer matrix relation dimer disorder} from the main text.

To present the transfer matrix, we first define
\bseq
\begin{align}
    D_n &= 1 + U_n^{ac} U_n^{db} e^{2i\epsilon t} - \cos(\dt)(U_n^{ac}+ U_n^{db})e^{i\epsilon t},\\
    E_n &= [U_n^{aa}U_n^{cc} - U_n^{ca}(U_n^{ac}+U_n^{db}) ]e^{i\epsilon t},\\
    F_n &= U_n^{ca} + U_n^{db}(U_n^{ac} U_n^{ca} - U_n^{aa}U_n^{cc} )e^{2i\epsilon t},\\
    G_n &= U_n^{bd} + U_n^{ac}(U_n^{bd}U_n^{db}-U_n^{bb}U_n^{dd})e^{2i\epsilon t} +\cos(\dt) [U_{n}^{bb}U_n^{dd} - U_n^{bd}(U_{n}^{ac} +U_n^{db} ) ]e^{i\epsilon t}. 
\end{align}
\eseq
Then the transfer matrix is
\beq
    \mathcal{M}_{n,\text{dimer}} = 
    \bpmat
        \mathcal{M}_{n,\text{dimer}}^{aa} & \mathcal{M}_{n,\text{dimer}}^{ad}\\
        \mathcal{M}_{n,\text{dimer}}^{da} & \mathcal{M}_{n,\text{dimer}}^{dd}
    \epmat,
\eeq
with the following matrix elements:
\bseq
\begin{align}
    \mathcal{M}_{n,\text{dimer}}^{aa} &= \frac{[E_n + \sec(\dt)F_n]}{D_n}e^{i\epsilon t},\\
    \mathcal{M}_{n,\text{dimer}}^{ad} &= i \{ \tan(\dt) [ 1+ (U_n^{ac} U_n^{db} + U_n^{cc} U_n^{dd}) e^{2i\epsilon t} ] - \sin(\dt) (U_n^{ac}+U_n^{db})e^{i\epsilon t} \}/D_n,\\
    \mathcal{M}_{n,\text{dimer}}^{da} &= -i \frac{\sin(\dt)E_n + \tan(\dt) F_n}{D_n G_{n+1}} \notag\\
    &\qquad \qquad \times \left[1+ (U_{n+1}^{aa}U_{n+1}^{bb} + U_{n+1}^{ac}U_{n+1}^{db} )e^{2i\epsilon t} -\cos(\dt)(U_{n+1}^{ac}+U_{n+1}^{db})e^{i\epsilon t}\right],\\
    \mathcal{M}_{n,\text{dimer}}^{dd}&= \biggr\{\sec (\text{$\Delta $t}) \left(1+e^{2 i t \epsilon } U_{n}^{ac} U_{n}^{db}\right) \left(1+e^{2 i t \epsilon } U_{n+1}^{ac} U_{n+1}^{db}\right)\notag\\
    &\times \cos (\text{$\Delta $t}) e^{ i t \epsilon } \biggr[\left(U_{n}^{ac}+U_{n}^{db}\right) \left(U_{n+1}^{ac}+U_{n+1}^{db}\right)+\sin ^2(\text{$\Delta $t}) e^{2 i t \epsilon } U_{n+1}^{aa} U_{n+1}^{bb} U_{n}^{cc} U_{n}^{dd}\biggr]\notag \\
    & -e^{2 i t \epsilon } U_{n+1}^{ac} U_{n}^{db} U_{n+1}^{db}-U_{n+1}^{ac}-U_{n+1}^{db}-U_{n}^{db}\notag\\
    & \ -U_{n}^{ac} \left\{1+e^{2 i t \epsilon } \left[U_{n+1}^{ac} U_{n+1}^{db}+U_{n}^{db} \left(U_{n+1}^{ac}+U_{n+1}^{db}\right)\right]\right\} \notag\\
    & \ - \sin ^2(\text{$\Delta $t}) e^{2 i t \epsilon } \left[U_{n+1}^{aa} U_{n+1}^{bb} \left(U_{n}^{ac}+U_{n}^{db}\right)+U_{n}^{cc} U_{n}^{dd} \left(U_{n+1}^{ac}+U_{n+1}^{db}\right)\right]\notag \\
    & \ + \sin (\text{$\Delta $t}) \tan (\text{$\Delta $t}) e^{i t \epsilon } \biggr[ U_{n+1}^{aa} U_{n+1}^{bb} \left(1+e^{2 i t \epsilon } U_{n}^{ac} U_{n}^{db}\right) +U_{n}^{cc} U_{n}^{dd} \left(1+e^{2 i t \epsilon } U_{n+1}^{ac} U_{n+1}^{db}\right)\biggr]\notag \\
    & + \sin ^3(\text{$\Delta $t}) \tan (\text{$\Delta $t}) +e^{3 i t \epsilon } U_{n+1}^{aa} U_{n+1}^{bb} U_{n}^{cc} U_{n}^{dd} \biggr\} / (D_n G_{n+1}).
\end{align}
\eseq

\section{Disorder remains short-ranged under loop decomposition}\label{sec:Disorder remains short-ranged under loop decomposition}

An important part of our scattering argument in Sec. \ref{sec:Scattering argument for the universal exponent} is the assumption that the disorder in the loop drive is short-ranged.  In this Appendix, we do some basic numerical checks to verify that this is a reasonable assumption, taking as a starting point that the disorder in the original (nonloop) drive is short-ranged.

Roughly speaking, the scenario we wish to rule out is the following.  If the original drive is a finite-sized disordered sample (see below), one could question if the loop drive is also described by a finite-sized disordered sample.  If instead it turns out that the loop decomposition spreads the disorder through the entire system, this would seem to make a scattering treatment impossible. 

More precisely, instead of taking the loop drive to have the form of Eqs. \eqref{eq:H_for_scattering} and \eqref{eq:Hdisorder}, we assume that the original drive takes that form, i.e.,
\beq
    \mathcal{H}(t) = \mathcal{H}_\text{clean}(t)+ \mathcal{H}_\text{disorder}(t),\label{eq:H_for_scattering_non-loop}
\eeq
where
\beq
    \mathcal{H}_\text{disorder}(t) = \sum_n \Theta_n'(t),\label{eq:Hdisorder_non-loop}
\eeq
and each operator $\Theta_n'(t)$ is exponentially localized near the site $n$.  We then do the loop decomposition, as described in Sec. \ref{sec:Preliminary discussion}, on both $\mathcal{H}(t)$ and $\mathcal{H}_\text{clean}(t)$; this yields loop drives $U(t)$ and $U_\text{clean}(t)$.  To match the main text setup exactly, we would need to find that the loop Hamiltonian takes the form \eqref{eq:H_for_scattering} and \eqref{eq:Hdisorder}.  However, this requirement may be too stringent.

A weaker condition that should still suffice for our purposes is that $U(t)$ should be suitable for a scattering treatment.  In particular, if we define a disordered sample in the usual way, by setting $\Theta_n'(t)= 0$ for all $n \le 0$ and $n \ge N+1$, we would then like to find that $U_{nn'}(t) \to U_{\text{clean},nn'}(t)$ far from the disordered region.

We have verified this condition numerically in a particular example of a nonloop drive.  We consider the SSH-type drive from the main text, generalized to have arbitrary coefficients in front of the trivial and topological SSH Hamiltonians.  Thus, we define
\bseq
\begin{align} 
    \mathcal{H}_{1,\text{clean}} &= h_1\sum_n (\ket{n,A}\bra{n,B} + \text{H.c.}),\label{eq:H1_non-loop} \\
    \mathcal{H}_2 &= \mathcal{H}_{2,\text{clean}} + \mathcal{H}_{2,d},\\
    \mathcal{H}_{2,\text{clean}} &= - h_2\sum_n (\ket{n+1,A}\bra{n,B} + \text{H.c.}),\label{eq:H2_non-loop}\\
    \mathcal{H}(t) &=
    \begin{cases}
      \mathcal{H}_{1,\text{clean}} &  0\le t < \frac{\td}{4}\\
      \mathcal{H}_2 &  \frac{\td}{4} \le t\le   \frac{3\td}{4}\\
      \mathcal{H}_{1,\text{clean}} &  \frac{3\td}{4}<t\le  \td,
  \end{cases} \label{eq:model_drive_non-loop}
\end{align}
\eseq
with $\mathcal{H}_\text{clean}(t)$ defined similarly (by setting $\mathcal{H}_{2,d}=0$).  For $h_1=h_2=2\pi/\td$, $\mathcal{H}_\text{clean}(t)$ is the same as the SSH-type drive from the main text; more generally, the drive has chiral symmetry but is not a loop.  We have assumed for definiteness that the disorder is only in $\mathcal{H}_2$.  We further specialize to the disorder being a single defect at the bond connecting two adjacent sites $n_0$ and $n_0+1$:
\beq
    \mathcal{H}_{2,d} = h_d (\ket{n_0+1,A}\bra{n_0,B} + \text{H.c.} ).
\eeq

Putting the system on a ring of $L$ sites, we do exact diagonalization to calculate the loop drive $U(t)$ near the midpoint using Eq. \eqref{eq:Uloop(t) near midpoint} and $H_F = (i/\td)\ln \mathcal{U}(\td)$.  To test if the defect spreads under the loop decomposition, we examine matrix elements of the operator
\beq
    \Delta U(t) \equiv U(t) - U_\text{clean}(t).
\eeq
We first consider the quantity
\beq
    \text{max}\{ \abs{\Delta U_{n'\alpha',n\alpha}(t)} \}_{1\le n' \le L} \qquad (\alpha,\alpha' \in \{A,B\})\label{eq:max}
\eeq
as a function of $n$, finding that it is peaked at the defect (i.e., at $n=n_0$ or $n_0+1$).  We then fix $n$ at the peak value and confirm that $\abs{\Delta U_{n'\alpha',n\alpha}(t)}$ decays rapidly with increasing $|n'-n|$ (Fig. \ref{fig:loop_decomposition_test}).
\begin{figure}[htb]
\centering
    \includegraphics[width=.6\columnwidth]{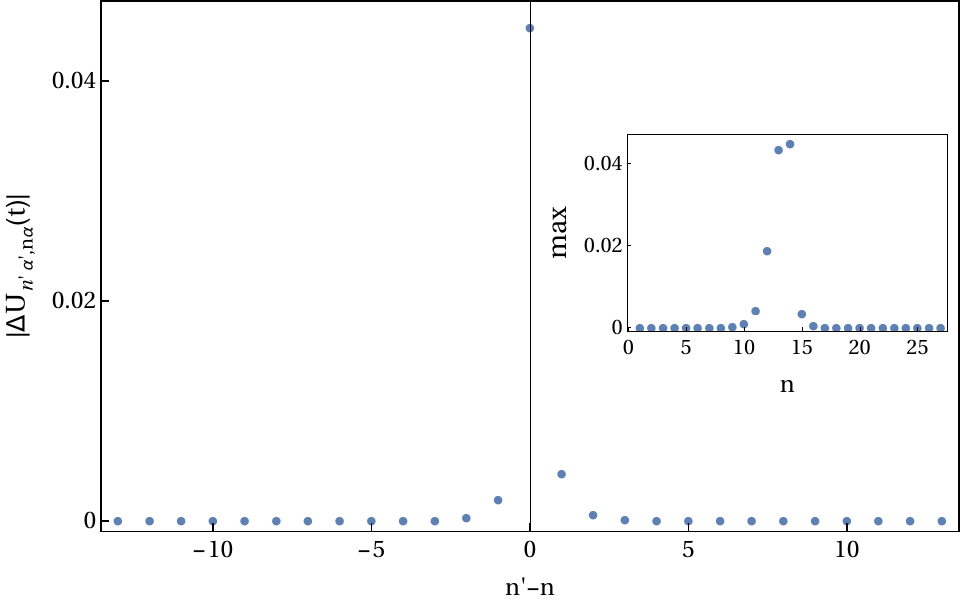}
    \caption{Numerical check that localized disorder does not spread under the loop decomposition.  We consider the model described in the main text with system size $L=27$, parameters $h_1=0.9, h_2=1.62, \td=2\pi, t= \td/2 + 0.01, h_d= 0.2$, and the defect site $n_0 =13$.  We fix $\alpha=\alpha'=A$, but similar results were obtained for other values.  In the main plot, we set $n=14$.  The inset plots the quantity \eqref{eq:max}, which is maximal at $n=14$.   }\label{fig:loop_decomposition_test}
\end{figure}

We obtained similar results in the case of a phase defect, where we define a chiral-symmetric nonloop drive in the same way as in Eq. \eqref{eq:disordered unitary one-sided}.  The disordered unitary $U_d$ is defined by setting all phases in Eq. \eqref{eq:Ud phase disorder} to zero except a single $\phi_{n_0,A}$.  In this case, we again found that the loop drive differs from the clean loop drive only near the defect site $n_0$.

\section{Reflection amplitude starts at linear order}\label{sec:Reflection amplitude starts at linear order}

Consider the scattering problem as defined in Sec. \ref{sec:Scattering argument for the universal exponent}, with a disordered region of size $N$ connected to clean leads, at some arbitrary $\dt$.  In this section, we use the notation
\bseq
\begin{align}
    U_0 &= U_\text{clean}(\td / 2),\\
    \widetilde{U} &= U(\td / 2),\\
    U&= U(\td / 2 + \dt ),
\end{align}
\eseq
and we also write
\bseq
\begin{align}
    \widetilde{U} &= U_0 + \widetilde{V},\\
    U&= U_0 + \widetilde{V} + V,
\end{align}
\eseq
where the disorder terms $V$ and $\widetilde{V}$ are exponentially localized to the sample region (the sites $1,\dots, N$).

Our task is to show the following: If the scattering problem for $\widetilde{U}$ has zero reflection, then the scattering problem for $U$ has reflection that generically starts at linear order in $V$.  This suffices because $V$ generically starts at linear order in $\dt$ [see Eq. \eqref{eq:generic expansion of U(t) near midpoint}].

We now proceed using the general scattering framework developed in Ref. \cite{BisioScattering2021} for discrete-time problems.  Consider an eigenstate $\ket{\Psi}$ of $U_0$ with some definite momentum (either right-moving or left-moving) and eigenvalue $e^{-i\omega}$.  The corresponding scattering ``in'' states for $\tilde{U}$ and $U$ satisfy the appropriate Lippmann-Schwinger equation:
\bseq
\begin{align}
    |\widetilde{\Psi}_\text{in}\rangle &= \ket{\Psi} + G_0 \widetilde{V}  |\widetilde{\Psi}_\text{in}\rangle,\label{eq:LS eqn for Utilde}\\
    \ket{\Psi_\text{in}} &= \ket{\Psi} + G_0(\widetilde{V} +V) \ket{\Psi_\text{in}}\label{eq:LS eqn for U},
\end{align}
\eseq
where $G_0 =(e^{-i(\omega + i 0^+)}- U_0)^{-1}$ is the free Green's function \cite{BisioScattering2021}.  Indeed, Eqs. \eqref{eq:LS eqn for Utilde} and \eqref{eq:LS eqn for U} follow from Eq. (S.3) from the Supplemental Material of Ref. \cite{BisioScattering2021} and the relation $\ket{\Psi_\text{in}}= \Omega_+ \ket{\Psi}$ (see also their footnote 53).

A similar equation relating the two scattering states is then readily obtained:
\beq
    \ket{\Psi_\text{in}}=  |\widetilde{\Psi}_\text{in}\rangle + \widetilde{G}V\ket{\Psi_\text{in}},
\eeq
where $\widetilde{G}= (e^{-i(\omega + i0^+)} - \widetilde{U})^{-1}$ is the Green's function of $\tilde{U}$.  Expanding in $V$, we get
\beq
    \ket{\Psi_\text{in}}=  |\widetilde{\Psi}_\text{in}\rangle + \widetilde{G}V|\widetilde{\Psi}_\text{in}\rangle + O(V^2).
\eeq
By assumption, $|\widetilde{\Psi}\rangle$ has zero reflection, i.e., its wavefunction agrees with that of $\ket{\Psi}$ for $n\to -\infty$ (assuming for definiteness that $\ket{\Psi}$ is right-moving).  The term linear in $V$ generically has nonvanishing amplitude for $n\to -\infty$, thus yielding the reflection amplitude ($r_{1\dots N}$) starting at linear order in $V$.

\end{widetext}

\bibliography{ms}

\end{document}